\documentclass[aip,pop,amsmath,amssymb,showpacs,reprint,floatfix,lengthcheck]{revtex4-1}
\usepackage[colorlinks,linkcolor=green,citecolor=green,bookmarks,pdfstartview=FitH]{hyperref}
\usepackage{graphicx}
\usepackage{dcolumn}
\usepackage{bm}
\usepackage{url}

\begin{document}

\title{Kinetic particle simulations in a global toroidal geometry}
\author{S. De}
\affiliation{Department of Physics, Indian Institute of Science, Bangalore 560012, India}
\author{T. Singh}
\affiliation{Department of Physics, Indian Institute of Science, Bangalore 560012, India}
\author{A. Kuley}
\email{akuley@iisc.ac.in}
\affiliation{Department of Physics, Indian Institute of Science, Bangalore 560012, India}
\author{J. Bao}
\affiliation{Department of Physics and Astronomy, University of California Irvine, CA 92697,USA}
\author{Z. Lin}
\affiliation{Department of Physics and Astronomy, University of California Irvine, CA 92697,USA}
\author{G. Y. Sun}
\affiliation{College of Physical Science and Technology, Xiamen University, Xiamen 361005, China}
\author{S. Sharma}
\affiliation{Institute for Plasma Research, Bhat, Gandhinagar-382428, India}
\affiliation{Homi Bhabha National Institute, Anushaktinagar, Mumbai 400 094, India}
\author{A. Sen}
\affiliation{Institute for Plasma Research, Bhat, Gandhinagar-382428, India}

\date{\today}

\begin{abstract}
The gyrokinetic toroidal code (GTC) has been upgraded for global simulations by coupling the core and scrape-off layer (SOL) regions across the separatrix with field-aligned particle-grid interpolations. A fully kinetic particle pusher for high frequency waves (ion cyclotron frequency and beyond) and a guiding center pusher for low frequency waves have been implemented using cylindrical coordinates in a global toroidal geometry. The two integrators correctly capture the particle orbits and agree well with each other, conserving energy and canonical angular momentum. As a verification and application of this new capability, ion guiding center simulations have been carried out to study ion orbit losses at the edge of the DIII-D tokamak for single null magnetic separatrix discharges. The ion loss conditions are examined as a function of the pitch angle for cases without and with a radial electric field. The simulations show good agreement with past theoretical results and with experimentally observed feature in which high energy ions flow out along the ion drift orbits and then hit the divertor plates. A measure of the ion direct orbit loss fraction shows that the loss fraction increases with the ion energy for DIII-D in the initial velocity space. Finally, as a further verification of the capability of the new code, self-consistent simulations of zonal flows in the core region of the DIII-D tokamak were carried out. All DIII-D simulations were perfomed in the absence of turbulence. 

\end{abstract}
\maketitle
\section{Introduction}
\label{intro}
One of the important challenges in achieving a viable operating regime for ITER and future fusion reactors is associated with the nonlinear turbulent dynamics of the plasma in the scrape-off layer (SOL) \cite{loarte2007power}. The plasma characteristics in SOL can greatly affect the overall confinement properties of the device and also regulate the heat load to the tokamak wall. It can also influence the level of fusion ash, the impurity dynamics, sheath physics and plasma shaping effects. Furthermore, the SOL dynamics can degrade the current drive performance of radio frequency (RF) waves, through their impact on the density threshold conditions for the onset of parametric decay instabilities \cite{cesario2010current}. An in-depth undertanding of the mechanisms determining the width of the SOL layer remains an outstanding open problem. A study of the SOL plasma dynamics is challenging due to the multiple spatial and temporal scales associated with different energy sources (instabilities) in that region. Fluid simulation transport codes such as UEDGE \cite{wising1996simulation} and SOLPS \cite{rozhansky2009new} are normally used to simulate the SOL dynamics. These fluid codes use a set of fluid transport equations that are based on the Braginskii equations \cite{braginskii1965review}. However, the results show a number of discrepancies between experimental findings and fluid simulations, especially in the characteristics of the radial electric field, parallel ion flow, impurity radiation, etc. \cite{chankin2007discrepancy,erents2004comparison,wischmeier2011assessment}. It is believed that kinetic effects could be a significant contributor in the SOL to processes like ion orbit losses \cite{pan2014co}, X point losses \cite{chang2002x}, nonlocal turbulent transport \cite{omotani2013non}, plasma sheath dynamics\cite{wesson1995effect}, parametric decay instabilities \cite{cesario2010current,kuley2009stabilization,kuley2010parametric,kuley2010lower}, etc. To correctly model many of these effects, one requires a kinetic approach that covers the closed and open field line regions across the separatrix, and includes the realistic SOL physics and tokamak geometry. Due to the difficulty in the accessibility of diagnostics in the SOL region, such global kinetic simulations can help develop useful insights for predicting the plasma dynamics in that region for present and future reactors such as ITER and DEMO. A laudable effort in this direction has been the development of the massively parallel kinetic simulation code XGC-1 \cite{chang2009compressed} that takes an approach based on first-principles and has emerged as an efficient method for describing the complex physics of turbulent transport. Another widely used and successful tool, the gyrokinetic toroidal code (GTC) has undergone continuous development for the past two decades and has been applied to the study of plasma transport in the core region \cite{lin1998turbulent}. GTC is a well-benchmarked, first-principles code which has been extensively applied to the investigation of neoclassical transport \cite {lin1997neoclassical,dong2016effects}, microturbulence \cite {xie2017new,fulton2016gyrokinetic,xiao2009turbulent,holod2016effect,xiao2015gyrokinetic,zhang2012nonlinear}, mesoscale Alfv$\acute e$n eigenmodes  \cite{zhang2012nonlinear,spong2012verification,wang2013radial} excited by energetic particles, macroscopic MHD modes \cite{mcclenaghan2014verification,liu2014verification,bao2017conservative} (kink and tearing modes) and radio frequency (RF) waves \cite{kuley2013verification,bao2014particle,kuley2015verification,kuley2015nonlinear,bao2015global,bao2016electromagnetic,bao2016nonlinear} in the core region. However, the assumptions used in studying turbulence in the core region may not be valid in the SOL region.
GTC normally uses conventional magnetic flux coordinates, in which the equations of motion encounter a mathematical singularity of the metric on the magnetic separatrix surface. This is due to vanishing of the poloidal magnetic field near X-point as well as singular behaviors of safety factor and Jacobian of Boozer coordinates near separatrix. Recently, the GTC was extended to separately study instabilities in the SOL and core regions of a field reversed configuration (FRC) using Boozer coordinates. However, the code still did not have the capability to couple these two regions\cite{fulton2016gyrokinetic,fulton2016bgyrokinetic}. The difficulty lay in the discontinuity of the poloidal angle across the separatrix, in the Boozer coordinates.  This limitation restricted the code's usage to electrostatic simulations either in the core or the SOL region, with no cross-separatrix coupling.

In our present work we report a significant enhancement of the GTC code, called global toroidal code using X point (GTC-X) through the development of a new global nonlinear particle simulation model that couples the tokamak core and SOL regions. The code also provides a realistic treatment of the separatrix region through the Equilibrium Fitting (EFIT) \cite{lao1990equilibrium,ren2011high} and IPREQ\cite{Srinivasan} equilibrium data files
 generated by experimental discharges. A particular feature of GTC-X is the use of a cylindrical coordinate system for the advancement of the particle dynamics, which allows particle motion in arbitrary shaped flux surfaces including the magnetic separatrix and the magnetic X-point in the tokamak. Currently, GTC-X has both fully kinetic and guiding center particle dynamics, but XGC-1 has only guiding center particle dynamics. The long-term goal of GTC-X is to develop electromagnetic (EM) global simulation model for coupling core and  SOL using both guiding center and fully kinetic particle dynamics. GTC-X in its present form is an electrostatic code, which has been applied to simulate ion temperature gradient instability in the FRC geometry, which has no toroidal magnetic field \cite{Bao19}. 
 
As a first step in developing this nonlinear particle simulation model, we have developed a method of field-aligned particle-grid interpolations using an axisymmetric mesh in the cylindrical coordinates, which takes advantage of the smallest number of grid points in the direction of the magnetic field with high resolution in any given poloidal plane. These field-aligned particle-grid interpolations can achieve the same numerical efficiency as the field-aligned mesh in magnetic flux cooordinates, employed earlier in the global code GTC \cite{xiao2015gyrokinetic}, and in the flux-tube codes such as FENICA \cite{hariri2013flux} and GEM \cite{Sturdevant2017}. The gain in computational efficiency by using appropriate coordinates and computational mesh helps to optimize turbulence simulations of large devices like ITER and DEMO.  We have also developed a fully kinetic (FK) particle pusher to capture the effect of high frequency waves (ion cyclotron frequency and beyond) and
a guiding center (GC) pusher to describe the particle dynamics associated with low frequency waves (much smaller than the ion cyclotron frequency).

To test the effectiveness of these enhancements and appropriately benchmark the code, we have carried out ion guiding center simulations to study ion orbit losses at the edge of the DIII-D tokamak for single null magnetic separatrix discharges. Using model calculations, some analytic expressions of such losses have been presented by Miyamoto \cite{miyamoto1996direct} and have been used in the past to estimate the loss region in velocity space for JET, JT-60 and ITER. Stacey has introduced the effect of ion orbit loss and X point loss in his fluid calculations for the interpretation of fluid transport in the edge region \cite{stacey2013effect,stacey2011effect}. We apply the new simulation model and benchmark GTC-X against some of these past results. We have examined the ion orbit losses as a function of the pitch angle, both in the presence and absence of an electric field. The simulations show good agreement with past theoretical results and with many experimentally observed features. A measure of the ion direct orbit loss fraction shows that the loss fraction increases with the ion energy for DIII-D in the initial velocity space.

As a testimony to the capability of GTC-X to reproduce the existing physical phenomena, we carried out self-consistent simulations of zonal flows in the core region of the DIII-D tokamak. The collisionless damping of the zonal electric field to a non-zero steady-state value, verifies the famous theory of Rosenbluth and Hinton \cite{Rosenbluth} on the collisionless damping of zonal flows.

The paper is organized as follows. Section \ref{equ_cord} contains a detailed description of the coordinate system and the representation of the equilibrium and fluctuating field quantities in the code. Section \ref{mesh} describes the computational mesh used in the simulations. The physics model explaining the dynamics of the particles is described in section \ref{phys_model} followed by our simulation results for the ion orbit losses in section \ref{ion_orbit_loss}. The verification of the damping of zonal flows is given in Section \ref{zonal_section}. Section \ref{disc} provides a brief summary and some concluding discussions.


 \section{Representation of equilibrium and coordinate system}
 \label{equ_cord}
In describing the collective dynamics of a tokamak plasma, it is convenient to represent the physical variables in terms of equilibrium and fluctuating quantities. The equilibrium is
described by the Grad-Shafranov equation. The fluctuations representing wave activity contribute to plasma transport. The input equilibrium magnetic field configuration for GTC-X can be generated in any of the equilibrium solvers, EFIT or IPREQ, and is normally expressed as a function of the magnetic flux function. In the MHD approximation (ignoring guiding center orbit effects) the plasma profiles of temperature, density, current etc. are also functions of the flux function.
In the GTC-X version we use cylindrical coordinates $(R,\zeta,Z)$ where $R$ is the distance from the geometry axis,  $\zeta$  is the
toroidal angle and $Z$ is the direction of the torus symmetry axis. The  representation of the magnetic field  
for an axisymmetric system is then given by,

\begin{figure*}[htbp]
\centering
\includegraphics[width=\textwidth]{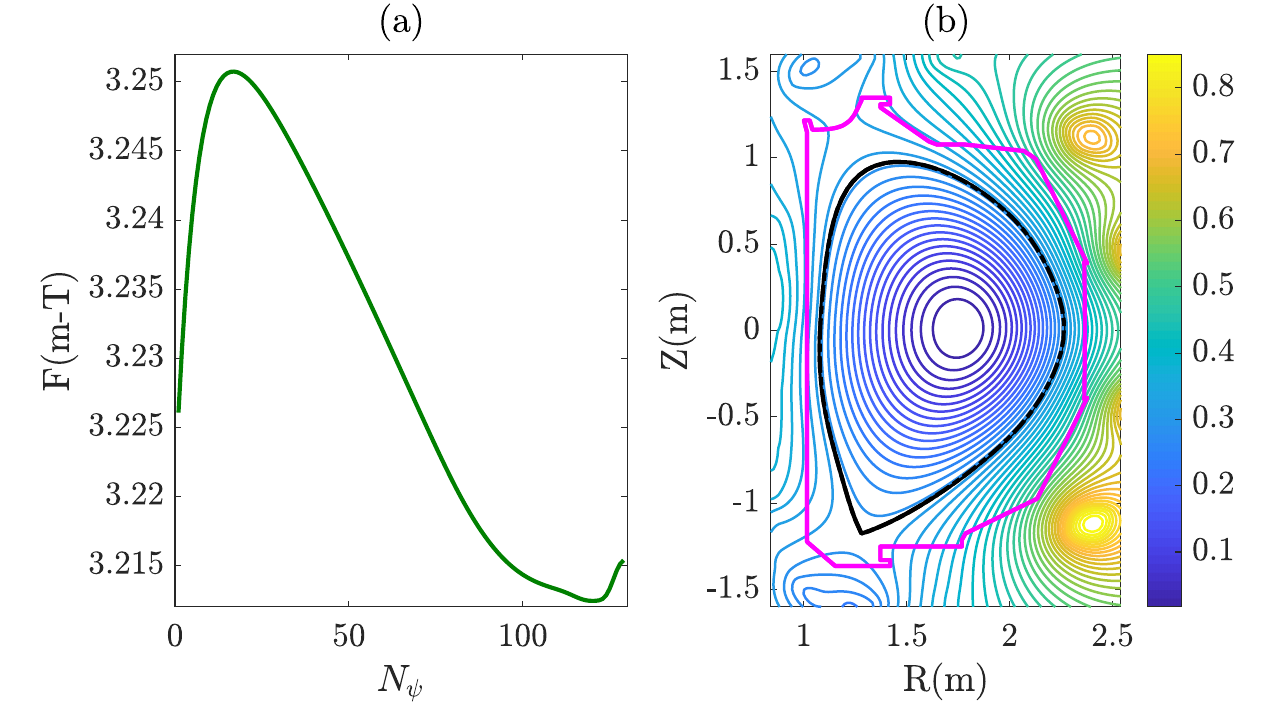}
\caption{\label{fig:Eq_profile}(a) Poloidal current function  $F(\psi)$ in  m-T on uniform flux grid, and
(b) poloidal flux function in web/rad on rectangular (R,Z) grid points for DIII-D shot $\#$158103 at 3050 ms \cite{PhysRevLett.114.105002}. The magnitude of the flux function is indicated by color. Last closed flux surface and limiter points are represented by black and 
magenta line, respectively.}
\end{figure*}

\begin{figure*}[htbp]
\centering
\includegraphics[width=\textwidth]{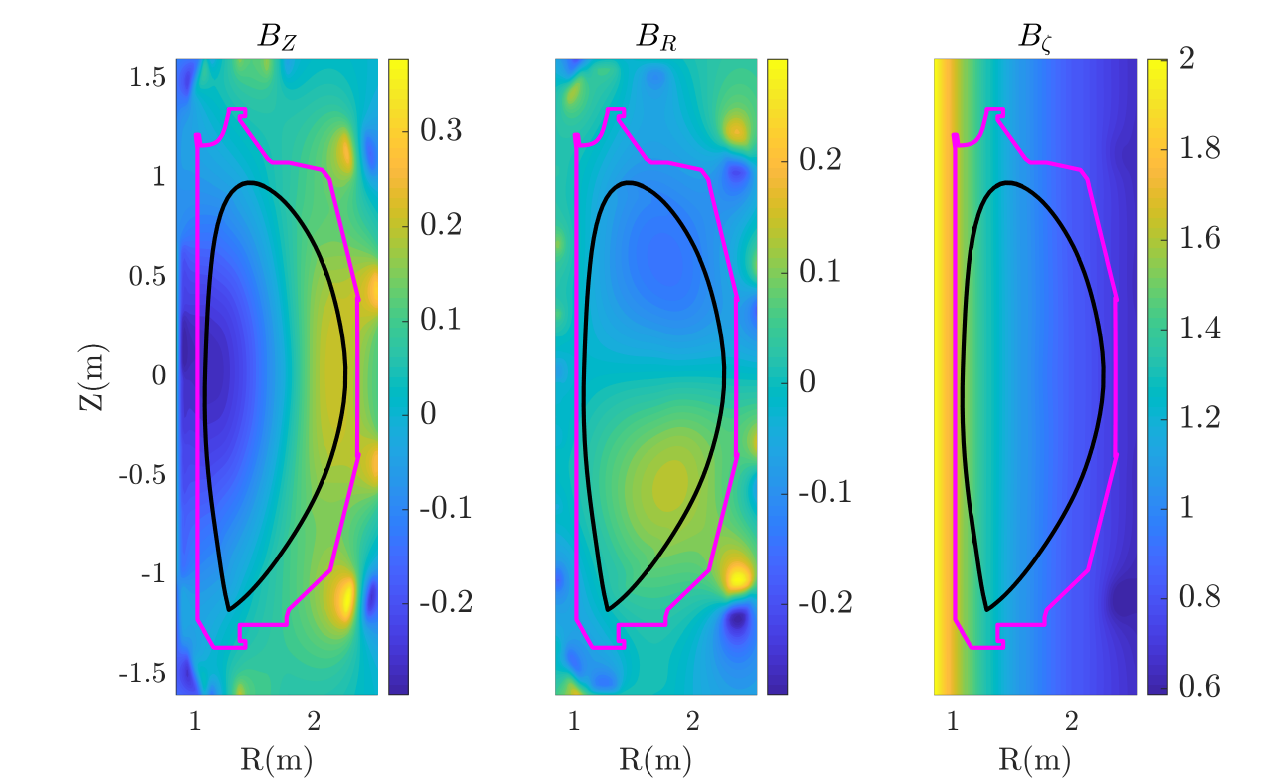}
\caption{Components of magnetic field (a) $B_Z$,
(b) $B_R$ and (c) $B_\zeta$ for DIII-D shot $\#$158103 at 3050 ms \cite{PhysRevLett.114.105002}.
The magnitude of the magnetic field components are indicated by color.
 Last closed flux surface and limiter points are represented by black and 
magneta line, respectively.}
\label{fig:Magneticfieldcomponents}
\end{figure*}

\begin{equation}
\vec B=\nabla\psi(R,Z)\times \nabla\zeta+\frac{F(\psi)}{R}\hat\zeta,
\label{eq:Eq_Bfield}
\end{equation}
where $\psi(R,Z)$ is the poloidal flux function that labels the magnetic surfaces for both closed and open field lines [cf. Fig~\ref{fig:Eq_profile}(b)] and $F(\psi)$ is the poloidal current function [cf. Fig~\ref{fig:Eq_profile}(a)], which 
provides the components of magnetic field in cylindrical coordinates as:
\begin{equation}
B_R=-\frac{1}{R}\frac{\partial\psi}{\partial Z},\quad B_Z=\frac{1}{R}\frac{\partial\psi}{\partial R},\quad B_\zeta=\frac{F(\psi)}{R},
\label{eq:magneticfieldcomponents}
\end{equation} 
The Jacobian for this system can be written as 
\begin{equation}
J^{-1}=\nabla R\cdot(\nabla\zeta\times\nabla Z).
\label{eq:Jacobian}
\end{equation}

The cylindrical toroidal coordinate system is related to the standard Cartesian system as follows
\begin{eqnarray}
 x=R\cos\zeta,\nonumber\\
 y=R\sin\zeta,\nonumber\\
 z= Z.
 \label{eq:cartesian}
\end{eqnarray}
By defining contravariant basis vectors $\vec{e}^R=\nabla R$, $\vec{e}^\zeta=\nabla\zeta$, 
$\vec{e}^Z=\nabla Z$ the 
velocity and the electric field can be written as
\begin{equation}
 \vec{v}=v^R\vec{e}_R+v^{\zeta}\vec{e}_\zeta+v^Z\vec{e}_Z,
\end{equation}

 \begin{equation}
 \vec{\Bbb E}=-\nabla\phi=-\biggl[\frac{\partial\phi}{\partial R}\nabla R+\frac{\partial\phi}{\partial\zeta}\nabla\zeta+
 \frac{\partial\phi}{\partial Z}\nabla Z\biggr],
\end{equation}
where
\begin{eqnarray}
 v^R=\dot{R},\quad v^{\zeta}=\dot{\zeta}, \quad v^Z=\dot{Z},\nonumber\\
 \vec{e}_R=\cos\zeta\hat{x}+\sin\zeta\hat{y},\nonumber\\
 \vec{e}_\zeta=-R\sin\zeta\hat{x}+R\cos\zeta\hat{y},\nonumber\\
 \vec{e}_Z=\hat{z}.
\end{eqnarray}

The equilibrium inputs from EFIT only provide equilibrium quantities on a coarse mesh, which usually contains a few tens of grid points in the $R$ and $Z$ directions. However, the micro-scale turbulence demands much denser grid points in $R$ and $Z$ directions. Therefore, it is necessary to map the equilibrium mesh to a dense computational mesh, in order to achieve sufficient numerical accuracy. 
For a 1D function $f(\psi)$, such as the poloidal current function $F(\psi)$, we can use the following B-spline representation \cite{xiao2015gyrokinetic}
\begin{equation}
f(\psi)=f(1,i)+f(2,i)\Delta\psi+f(3,i)\Delta\psi^2,
\label{eq:1dspline}
\end{equation}
where $f(2,i)$ and $f(3,i)$ are coefficients related to the first and second order differential in $\psi$ direction, which are calculated using finite difference method on a spline mesh. In our simulations we have calculated the grid size in $\psi$ using the following three steps:
\begin{equation}
\Delta\psi=\frac{\psi_{Lim}}{N_\psi-1}\Rightarrow N_{sep}=\texttt{integer}\biggl[\frac{\psi_{sep}}{\Delta\psi}\biggr]
\Rightarrow\Delta\psi=\frac{\psi_{sep}}{N_{sep}},
\end{equation}
where $\psi_{Lim}$ is the poloidal flux function at the limiter point, $N_\psi$ is the number of grid points in $\psi$, $N_{sep}$ is the grid point number at the separatix, and $\psi_{sep}$ is the poloidal flux function at the separatrix. This calculation will provide an accurate value of $\psi$ at the separatrix in the simulation. However, it will give some minor differences in calculating the $\psi$ at the limiter points.

A 2D function $f(R,Z)=\sum_n \Bbb F_n(R)\Bbb G_n(Z)$ can be expressed as 
\begin{eqnarray}
f(R,Z)=f(1,i,j)+f(2,i,j)\Delta R+f(3,i,j)\Delta R^2+\quad\nonumber\\
f(4,i,j)\Delta Z+f(5,i,j)\Delta R\Delta Z+f(6,i,j)\Delta Z\Delta R^2+\nonumber\\
f(7,i,j)\Delta Z^2+f(8,i,j)\Delta R\Delta Z^2+f(9,i,j)\Delta R^2\Delta Z^2,\nonumber\\
\label{eq:2Dspline}
\end{eqnarray}
where $\Delta R=R_{i+1}-R_i$, $i=1,2,3,\cdots,nR$, and $\Delta Z=Z_{j+1}-Z_j$, $j=1,2,3,\cdots,nZ$. Eq.~(\ref{eq:2Dspline}) is derived by using the 1D B-spline functions of $\Bbb F_n(R)=\Bbb F_n(1,i)+\Bbb F_n(2,i)\Delta R+\Bbb F_n(3,i)\Delta R^2$ and $\Bbb G_n(Z)=\Bbb G_n(1,j)+\Bbb G_n(2,j)\Delta Z+\Bbb G_n(3,j)\Delta Z^2$. The spline coefficients $f(1:9,i,j)$ are calculated from the spline coefficients $\Bbb F_n(1:3,i)$ and $\Bbb G_n(1:3,j)$. Fig.~\ref{fig:Eq_profile}(a) and Fig.~\ref{fig:Eq_profile}(b) represent the equilibrium poloidal current function on a uniform flux grid and poloidal flux function on the rectangular R-Z grid respectively, for DIII-D shot $\#$158103 at 3050 ms \cite{PhysRevLett.114.105002}. In GTC-X we use these two functions in Eq.(2) to calculate the magnetic field components for DIII-D [cf. Fig.~\ref{fig:Magneticfieldcomponents}]. In the particle pusher, we interpolate these field quantities at the particle position using a 2D spline function, as described in Eq.~(\ref{eq:2Dspline}). The magnetic field is divergence-less if we use the expressions of $B_R, B_Z$, and $B_\zeta$ as given by Eq.(2). The equilibrium poloidal flux $\psi$ is numerically calculated by a 2D spline representations. Hence for this axisymmetric system, the $\nabla\cdot \vec B=0$ is guaranteed numerically in the simulation.


\section{Field aligned mesh for fluctuating quantities}
\label{mesh}
The field aligned mesh has maximum numerical efficiency and accuracy to address the nonlinear physics of drift wave turbulence. Because the particle moves much faster in the direction parallel to the magnetic field than when it drifts across the magnetic field, the parallel wave vector is usually much smaller than the perpendicular wave vector for the drift wave instabilities. Thus, one only requires a small number of grid points to resolve the parallel wave length, which greatly saves the computational costs and suppresses the numerical high $k_{||}$ modes efficiently. Furthermore, it helps to simplify the implementation of the field solver, since the field aligned mesh can exactly decouple the parallel and perpendicular directions. 

\subsection{Radial grid:} GTC code has in the past utilized the field aligned mesh in Boozer coordinates and has successfully applied it to gyrokinetic simulations of drift waves in the core region of toroidal plasmas \cite{lin1998turbulent}. In this work, we extend the original GTC field aligned mesh from Boozer coordinates in the core region to cylindrical coordinates in the whole domain across the separatrix. In order to create the field aligned mesh for the whole tokamak domain, we build equi-spaced radial grids (spacing of size $\Delta r$) on the outer mid-plane for each field line, which can be calculated as:

\begin{eqnarray}
\Delta r=\frac{1}{{N_{SOL}-1}}\biggl\{R_{LF}(\psi=\psi_1,Z=0)-\nonumber\\
R_{LF}(\psi=\psi_{sep},Z=0)\biggr\},
\end{eqnarray}
where $R_{LF}$ is the radial position as a function of poloidal flux and $Z$ on the low field side, and $N_{SOL}$ is the number of field lines in the SOL region. $\psi_1$ is the maximum value of the poloidal flux on the limiter points (plasma facing components). GTC-X also has the capability to use a nonuniform grid, with the grid size in the perpendicular direction correlated with the local gyro-radius. The poloidal flux at each field line of the simulation mesh in the SOL region is: 

\begin{eqnarray}
\psi_{SOL}(i_{SOL})=\psi\biggl(R_{LF}(\psi=\psi_{sep},Z=0)+\nonumber\\
(i_{SOL}-1)\times\Delta r,Z=0\biggr),
\end{eqnarray}
where $i_{SOL}\in[1,N_{SOL}]$. 

Then the field line number in the core region $N_{core}$ can be calculated as: 

\begin{equation}
N_{core}=\texttt{integer}\biggl[\frac{R_{LF}(\psi=\psi_{sep},Z=0)-R_0}{\Delta r}\biggr]
\end{equation}
where $R_0$ is the major radius. The poloidal flux at the innermost grids in the core region $\psi_0$ can be determined as:

\begin{equation}
\psi_{0}=\psi\biggl(R_{LF}(\psi=\psi_{sep},Z=0)-(N_{core}-1)\times\Delta r,Z=0\biggr),
\end{equation}
where $\psi_0$ is not necessarily equal to the poloidal flux value at the magnetic axis. The poloidal flux of each field line for the simulation mesh in the core region is: 
\begin{eqnarray}
\psi_{core}(i_{core})=\psi\biggl(R_{LF}(\psi=\psi_0,Z=0)+\nonumber\\
(i_{core}-1)\times\Delta r,Z=0\biggr),
\end{eqnarray}
where $i_{core}\in[1,N_{core}]$.

Similarly, the field line number in the private region (i.e., the space/dome below separatrix X-point, which has no connection of flux lines with main plasma) $N_{private}$ is calculated as:
\begin{eqnarray}
N_{private}=\texttt{Integer}\biggl[\frac{R_{LF}(\psi=\psi_{sep},Z=Z_0)}{\Delta r}-\nonumber\\
\frac{R_{LF}(\psi=\psi_{pmax},Z=Z_0)}{\Delta r}\biggr]
\end{eqnarray}
where $\psi_{pmax}$ is the poloidal flux at the innermost flux surface in the private region, and $Z_0$ is the minimum value in $Z$ direction. The poloidal flux at the innermost grids in the private region $\psi_{p0}$ can be determined as:
\begin{eqnarray}
\psi_{p0}=\psi\biggl(R_{LF}(\psi=\psi_{sep},Z=0)-\nonumber\\
(N_{private}-1)\times\Delta r,Z=Z_0\biggr),
\end{eqnarray}
where $\psi_{p0}$ is not necessarily equal to $\psi_{pmax}$. The poloidal flux at each field line of the simulation mesh in the private region is: 
\begin{eqnarray}
\psi_{private}(i_{private})=\psi\biggl(R_{LF}(\psi=\psi_{p0},Z=Z_0)+\nonumber\\
(i_{private}-1)\times\Delta r,Z=Z_0\biggr),
\end{eqnarray}
where $i_{private}\in[1,N_{private}]$. 

\subsection{Poloidal grid:}
After this, we trace each field line along the poloidal direction and calculate the length of the poloidal projection of each field line by using the unit magnetic vector as follows:
\begin{equation}
\frac{dR}{dS}=b_R,
\end{equation}
and
\begin{equation}
\frac{dZ}{dS}=b_Z,
\end{equation}
Here, $b_R=-(1/B_pR)(\partial\psi/\partial Z)$, $b_Z=(1/B_pR)(\partial\psi/\partial R)$ and $B_p$ is the poloidal magnetic field. The starting points for tracing the field lines are on the low field side of outer mid-plane ($Z=0$ plane) for the core region and, on the low field side of the $Z=Z_0$ plane for the SOL and the private regions. The step $\Delta S_t$ for tracing each field line is at least $10^{-5}$ smaller than the order of magnitude of the length of the poloidally projected field line, and we correct the new position for each tracing step by using Newton’s method, with the error of the order of floating precision, in order to ensure that it lies on the same flux surface,. All the advanced positions during the tracing of the poloidal field line are recorded as tracing grids. The outermost simulation boundary is determined by the limiter, as shown by the brown line. As a result, the tracing grids outside the limiter in the SOL and private region are removed by using ray casting algorithm \cite{wiki.org}. In the closed field line region, we calculate the length of the poloidal projection of the field line for each flux surface. However, in the open field line regions, we set the intersection points of the field lines and the limiter as the simulation boundary points and then calculate the length of the poloidal projection of the field line between each of the two intersection points on a single continuous field line.

The number of simulation grids for each continuous poloidal field line is calculated as,
\begin{equation}
N_i=\texttt{integer}(\frac{l_i}{\Delta S_{i0}}),
\end{equation}
where $l_i$ is the poloidal field line length of i-th field line and $\Delta S_{i0}$ is the approximate grid size along the poloidal direction. The exact poloidal grid size $\Delta S_i$ is then determined by $l_i$ and $N_i$ as:
\begin{equation}
\Delta S_i=\frac{l_i}{N_i-1}.
\end{equation}
The positions of the field aligned grid along poloidal direction can be derived from poloidal grid size $\Delta S_i$, the much smaller tracing step $\Delta S_t$ and the positions of the dense tracing grids.

In the cylindrical coordinate representation, we only create the field aligned mesh in the poloidal direction, and the meshes on different poloidal plane are identical (as shown in Fig.~\ref{fig:Orbit}) due to axisymmetry. In order to decouple the exact parallel and perpendicular directions and use a small number of poloidal planes, the particle-grid gathering and scattering operations can be done exactly along the parallel direction by interpolating the value on the identical field line.  
\section{Physics model for particle dynamics}
\label{phys_model}
The efficiency of particle simulation strongly depends on the advancement of the dynamical quantities. 
In the GTC-X we have developed particle pushers for both
fully kinetic particles as well as guiding center particles using cylindrical coordinates $(R,\zeta,Z)$ in a global toroidal geometry.
The models for the fully kinetic dynamics and guiding center dynamics
and, the numerical methods associated with the time advancement of the physical quantities
(particle position and guiding center) are described in the following sections.

\subsection{Fully kinetic particle dynamics}
Fully kinetic particle dynamics is described by the six dimensional Vlasov equation
 \begin{equation}
\biggl[\frac{\partial}{\partial t}+\vec{v}\cdot \nabla+\frac{q_c}{m}(\vec{\Bbb E}+
\vec{v}\times \vec{B})\cdot\frac{\partial}{\partial \vec{v}}\biggr]f_{FK}=0,
\end{equation}
where $f_{FK}$ is the fully kinetic particle distribution function, $q_c$ is the particle charge, and $m$ is the particle mass. The above equation governs the evolution of $f_{FK}$ in the fully kinetic approach.

The time evolution of the phase space coordinates of a single particle, in the presence of a self-consistent electromagnetic field, is governed by the Lorentz-force equation as follows:
\begin{equation}
 \frac{d}{dt}\vec{r}=\vec{v},\quad\quad
 \frac{d}{dt}\vec{v}=\frac{q_c}{m}\biggl[\vec{\Bbb E}+\vec{v}\times\vec{B}\biggr].
 \label{eq:Newton}
\end{equation}
In our simulation we compute the marker particle trajectory Eq.~(\ref{eq:Newton}) using the time centered Boris push
method {\cite{kuley2013verification,kuley2015verification,Boris,Birdsall,Tajima,wei2015method} as discussed in  section 4.2. The Lagrangian for the single 
particle motion in the cylindrical  coordinates is written as 
\begin{equation}
\mathcal{L}=\frac{m}{2}\biggl[\dot R^2+R^2\dot\zeta^2+\dot Z^2\biggr]+q_c\biggl[\dot R A_R+R\dot\zeta A_\zeta+\dot Z A_Z\biggr]-q_c\phi,
\label{eq:Lagrangian}
\end{equation} 
where $\vec A$ is the magnetic vector potential. Now the components of generalized momenta are 
\begin{eqnarray}
p_R=\frac{\partial\mathcal{L}}{\partial \dot R}=m\dot R+q_cA_R,\nonumber\\
p_\zeta=\frac{\partial\mathcal{L}}{\partial \dot\zeta}=mR^2\dot \zeta+q_cRA_\zeta,\nonumber\\
p_Z=\frac{\partial\mathcal{L}}{\partial \dot Z}=m\dot Z+q_cA_Z.
\label{eq:canonicalcomponentsfk}
\end{eqnarray}

In our simulation we use the poloidal flux function $\psi$, rather than the vector potential $\vec A$ to represent the equilibrium magnetic field components for calculating the toroidal canonical angular momentum. Two constants of motion for fully kinetic single-particle dynamics in cylindrical coordinates are defined by \cite{Sturdevant2017} :
\begin{itemize}
\item Kinetic Energy 
\begin{eqnarray}
\texttt{E}=(m/2)[(v^R)^2+(Rv^\zeta)^2+(v^Z)^2],\nonumber
\end{eqnarray}
\end{itemize}
\begin{itemize}
\item Toroidal angular momentum from Eq.~(\ref{eq:canonicalcomponentsfk}) and Eq.~(\ref{eq:Eq_Bfield})
\begin{eqnarray}
p_\zeta=mR^2v^\zeta+q_c\psi,\nonumber
\end{eqnarray}
since $RA_\zeta=\psi$
\end{itemize}

\begin{figure*}[htbp]
\centering
\includegraphics[width=14cm,height=21cm]{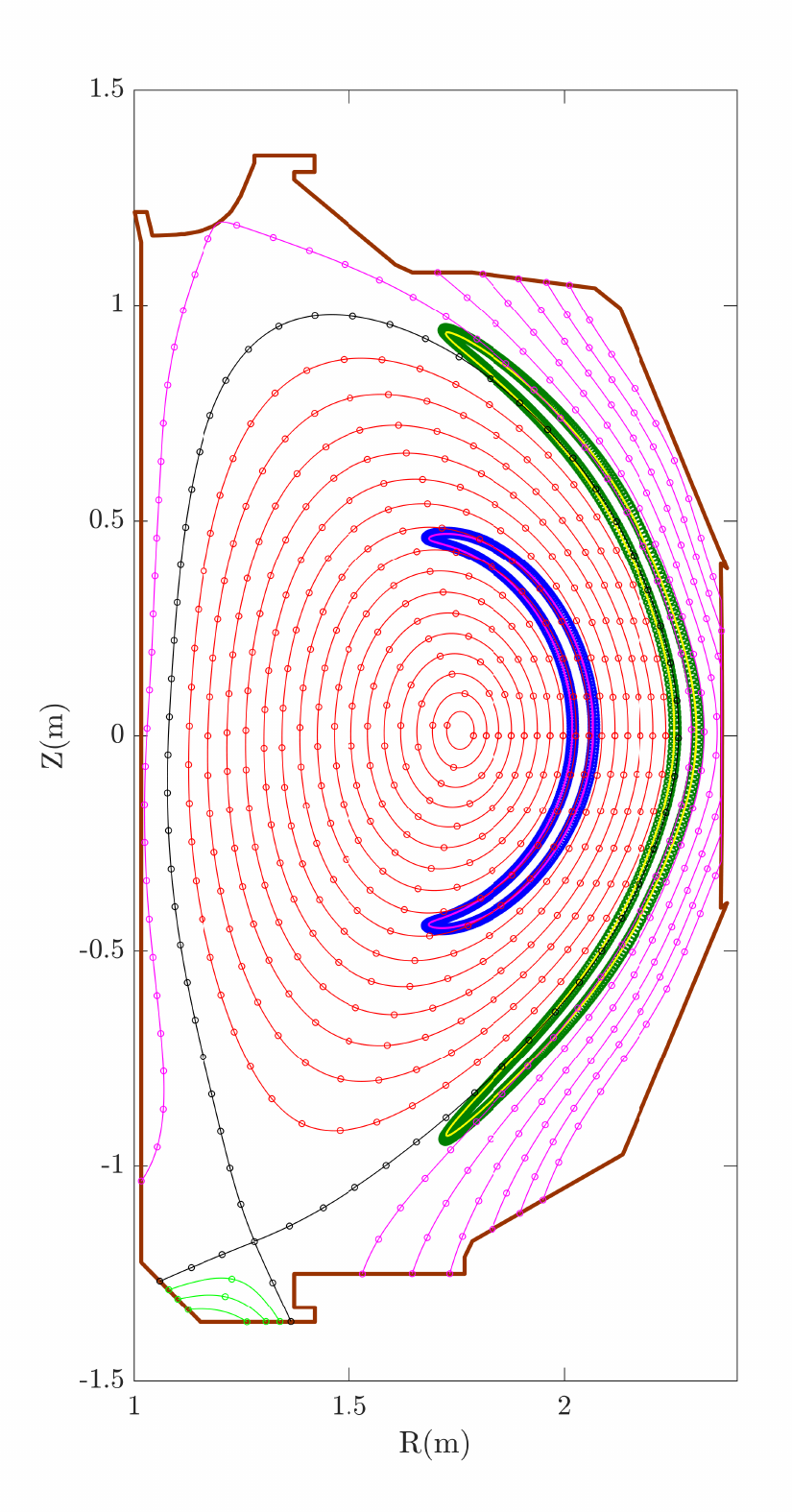}
\caption{GTC-X computational grids on a poloidal plane coupling core and SOL. Field aligned mesh at the core, separatrix, SOL, and private regions are represented by red, black, magenta, and green, respectively. Fully kinetic (blue and green) and guiding center (magneta and yellow) calculations of trapped particle orbits in the core  (51.66keV) and cross separatrix (59.42keV) for DIII-D shot $\#$158103 at 3050 ms \cite{PhysRevLett.114.105002}. Limiter points are represented by dark brown line.}
\label{fig:Orbit}
\end{figure*}
 \begin{figure}
 \begin{center}
 \includegraphics[scale=0.45]{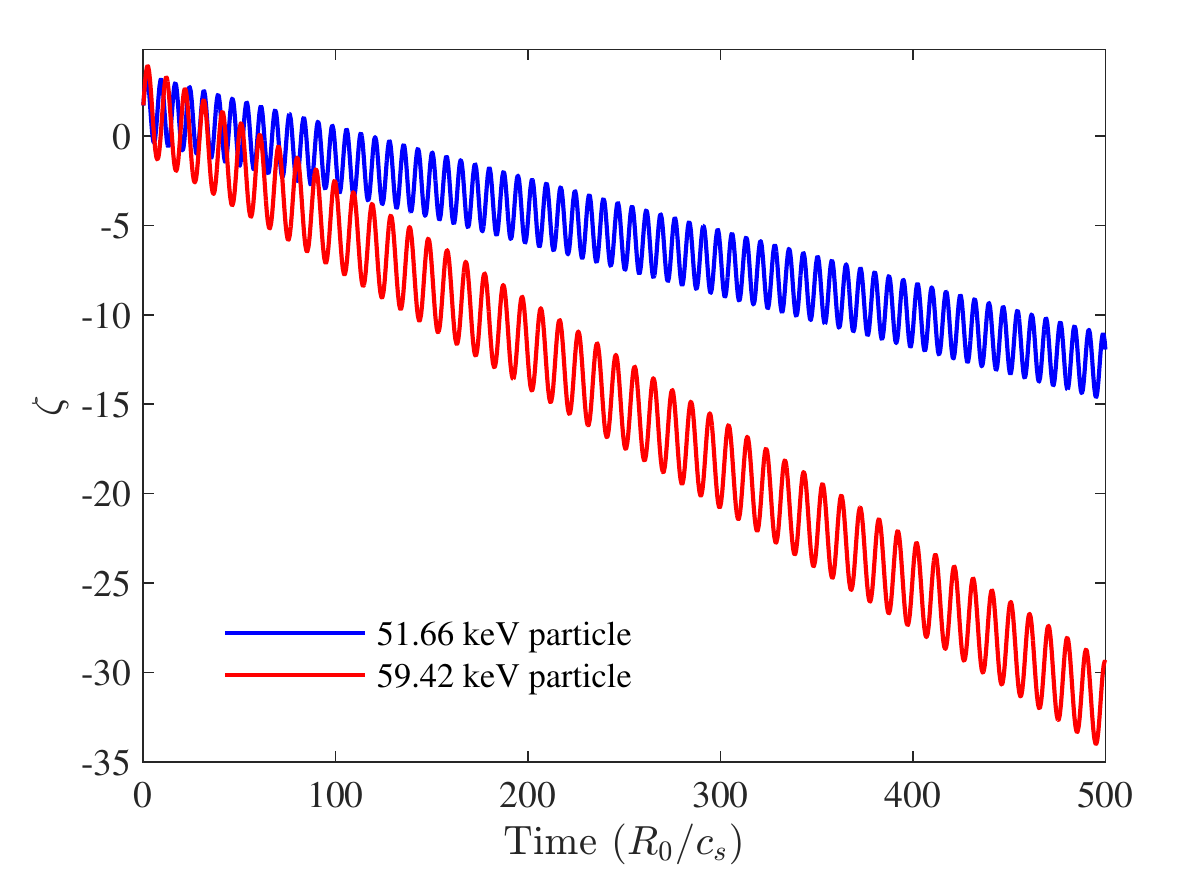}
 \caption{Variation of the toroidal angle $\zeta$ with time for the trapped particle orbits of Fig.~\ref{fig:Orbit}, in the core (blue) and the cross-separatrix (red) regions. The oscillations of $\zeta$ indicate the bounce motion and the gradual change in the mean level of the oscillations implies the toroidal precession of these bounce orbits.}
 \label{torprec}
 \end{center}
 \end{figure}
\subsection{Boris push for fully kinetic particle dynamics}
Boris scheme is the most 
widely used orbit integrator in explicit particle-in-cell (PIC)
simulation of plasmas. In this paper we have extended our Boris push
scheme in GTC from Boozer coordinates \cite{kuley2015verification} to cylindrical coordinates.
This scheme offers second order accuracy while requiring only one 
force (or field) evaluation per
step. The interplay between the PIC cycle and the Boris scheme is
schematically represented in Fig.2 of Ref.\cite{kuley2015verification}. 
At the beginning of each cycle the position of the particles and their
time centered velocity $\vec{v}(t-1/2)$
as well as the grid based electromagnetic fields $\vec{\Bbb E}(t),\vec{B}(t)$ are given.

At the first step, we add the first half of the electric field to update the velocity from $(t-1/2) $ to $t$ as follows:

\begin{equation}
\vec{u}(t)=\vec v(t-1/2)+\frac{q_c}{m}\frac{\Delta t}{2} \vec {\Bbb E}(t)
\end{equation}
One may write the components of velocity at particle position at $t$ as 
\begin{eqnarray}
 u^{\alpha-}(t)=\sum_{\beta=R,\zeta,Z}v^\beta(t-1/2)\vec{e}_\beta(t-1/2)\cdot\nabla\alpha(t)\nonumber\\
 +\frac{q_c}{m}\frac{\Delta t}{2}\vec{\Bbb E}(t)\cdot\nabla\alpha(t),\quad
 \label{eq:firsthalf}
\end{eqnarray}
where $\alpha=R,\zeta,Z$. For an orthogonal cylindrical system the above Eq.~(\ref{eq:firsthalf}) can be rewritten as as follows:
\begin{eqnarray}
u^{R-}(t)=A_1v^R(t-1/2)+B_1 R(1)v^\zeta(t-1/2)\nonumber\\
+\frac{q_c}{m}\frac{\Delta t}{2}\vec{\Bbb E}(t)\cdot\nabla R(t),\quad\nonumber\\
u^{\zeta-}(t)=R(2)\biggl[\frac{A_1R(1)}{g_{\zeta\zeta}(t)}v^\zeta(t-1/2)-\frac{B_1}{g_{\zeta\zeta}(t)}v^R(t-1/2)\biggr]\nonumber\\
\quad\quad\quad\quad+\frac{q_c}{m}\frac{\Delta t}{2}\vec{\Bbb E}(t)\cdot\nabla \zeta(t), \nonumber\\
u^{Z-}(t)=v^Z(t-1/2)+\frac{q_c}{m}\frac{\Delta t}{2}\vec{\Bbb E}(t)\cdot\nabla Z(t),\quad\quad\quad
\label{eq:velocitycomponents}
\end{eqnarray}
where $A_1=\cos(\zeta_2-\zeta_1),$ $B_1=\cos\zeta_1\sin\zeta_2-\sin\zeta_1\cos\zeta_2$, $\zeta_1=\zeta(t-1/2)$, 
$\zeta_2=\zeta(t)$, $R(1)=R(t-1/2)$ and $R(2)=R(t)$.

In the second step we consider the rotation of the velocity at time $(t)$. Rotated vector can be written as
\begin{equation}
 \vec{u}^+(t)=\vec{u}^-(t)+\vec{u}^-(t)\times\vec{s}(t)+[\vec{u}^-(t)\times\vec{T}(t)]\times\vec{s}(t),
\end{equation}
where $\vec{T}=(q_c\vec{B}/m)(\Delta t/2)$ and $\vec{s}=2\vec{T}/(1+T^2)$. The components of rotated vector become
\begin{eqnarray}
u^{R+}(t)=\biggl[1-PQ\biggl(B_Z^2+B_\zeta^2\biggr)\biggr]u^{R-}(t)\nonumber\\
+\biggl[PQB_RB_\zeta R+PB_Z\frac{\texttt{g}_{\zeta\zeta}}{J}\biggr]u^{\zeta-}(t)+\nonumber\\
 \biggl[PQB_RB_Z-PB_\zeta R \frac{\texttt{1}}{J}\biggr]u^{Z-}(t),
\end{eqnarray}

\begin{eqnarray}
u^{\zeta+}(t)=\biggl[1-PQ\biggl(B_R^2+B_Z^2\biggr)\biggr]u^{\zeta-}(t)\nonumber\\
+\biggl[PQB_RB_\zeta R \texttt{g}^{\zeta\zeta}-PB_Z\frac{\texttt{1}}{J}\biggr]u^{R-}(t)\nonumber\\
\biggl[PQB_ZB_\zeta R \texttt{g}^{\zeta\zeta}+PB_R\frac{\texttt{1}}{J}\biggr]u^{Z-}(t),
\end{eqnarray}
\begin{eqnarray}
u^{Z+}(t)=\biggl[1-PQ\biggl(B_R^2+B_\zeta^2\biggr)\biggr]u^{Z-}(t)\nonumber\\
+\biggl[PQB_ZB_\zeta R-PB_R\frac{\texttt{g}_{\zeta\zeta}}{J}\biggr]u^{\zeta-}(t)\nonumber\\
+\biggl[PQB_RB_Z+PRB_\zeta\frac{\texttt{1}}{J}\biggr]u^{R-}(t),
\end{eqnarray}
where  $Q=(q_c/m)(\Delta t/2)$, $P=2/(1+T^2)(q_c/m)(\Delta t/2)$, $g_{\zeta\zeta}=R^2$, and $J^2=det(\texttt{g}_{\alpha\beta})=R^2$.
In the third step, we
add the other half electric acceleration to the rotated vectors to obtain the velocity at time $(t+1/2)$
\begin{equation}
 u^\alpha(t+1/2)=u^{\alpha+}(t)+\frac{q_c}{m}\frac{\Delta t}{2}\vec{\Bbb E}(t)\cdot\nabla\alpha(t).
\end{equation}
To update the particle position we need to recover $\vec{v}(t+1/2)$, which can be done 
through the following transformation (cf. Fig.2 of Ref.\cite{kuley2015verification} dark purple arrow)
\begin{equation}
 v^\gamma(t+1/2)=\sum_{\alpha=R,\zeta,Z}u^{\alpha}(t+1/2)\vec{e}_\alpha(t)\cdot\nabla\gamma(t+1/2),
 \label{eq:finalvelocity}
\end{equation}
where $\gamma=R,\zeta,Z$. However, the basis vector $\nabla\gamma(t+1/2)$ is still unknown, since
$\gamma(t+1/2)$ does not exist in standard leap-frog scheme. Here we use a prediction for $\gamma(t+1/2)$ as
\begin{equation}
 \gamma(t+1/2)=\gamma(t)+u^\gamma(t+1/2)\frac{\Delta t}{2}.
\end{equation}
After we find the velocity at time $(t+1/2)$, we can update the particle position using the leap-frog scheme as
\begin{equation}
 \gamma(t+1)=\gamma(t)+v^\gamma(t+1/2)\Delta t.
\end{equation}
In Eq.~(\ref{eq:finalvelocity}) we have the dot-product of two basis vectors at different time steps. We have calculated this equation in the similar fashion as discussed in Eqs.~(\ref{eq:firsthalf})-~(\ref{eq:velocitycomponents}).

In this section, we have described the time advancement of the
dynamical quantities such as velocity and position of the particle using time centered approach. However, 
the self-consistent simulation requires update of particle
weight (representing perturbed distribution function), guiding center and electric field.  In our future simulations, we will use the second order Runge-Kutta (RK) method, 
to advance these quantities \cite{kuley2015verification,kuley2015nonlinear}.

\begin{figure*}[htbp]
\centering
\includegraphics[width=\textwidth]{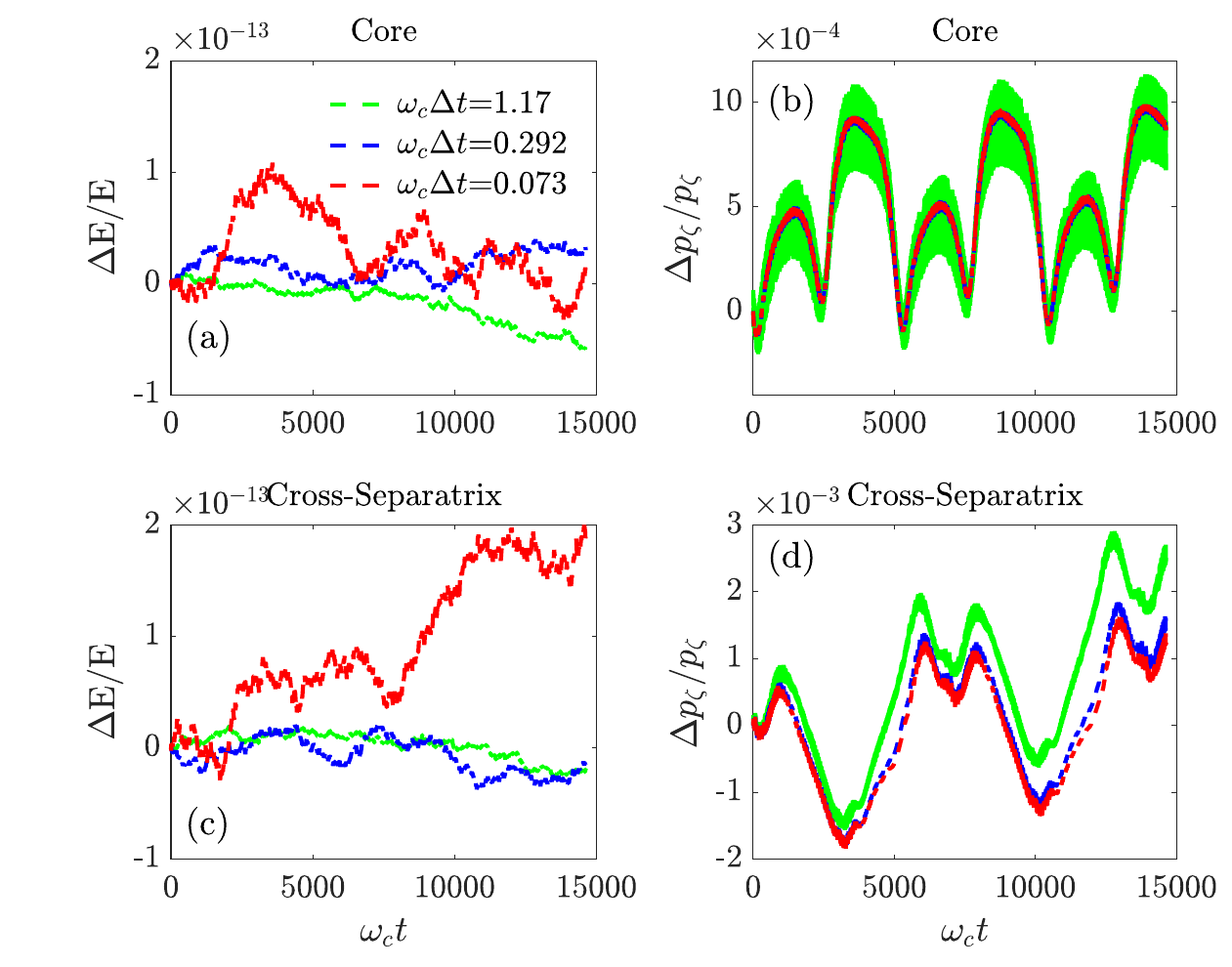}
\caption{Time step convergence of fully kinetic Boris integrator. Fig.(a), Fig.(c) represent the relative energy error (in the range of floating point cutoff error) and panel.(b), Fig.(d) show relative canonical angular momentum error for DIII-D geometry core and cross-separatrix region, respectively.}
\label{fig:FKconvergence}
\end{figure*}

\begin{figure*}[htbp]
\centering
\includegraphics[width=\textwidth]{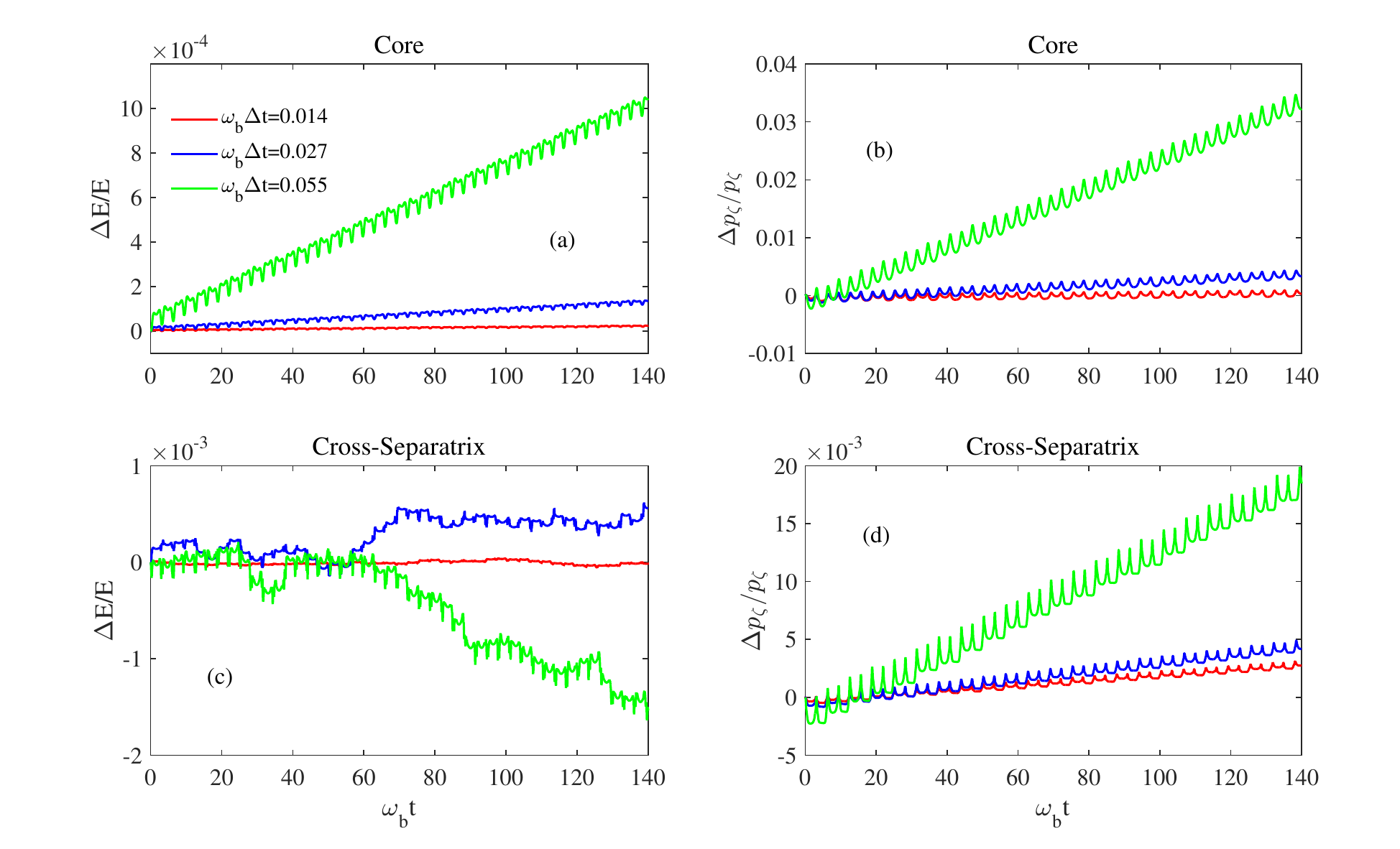}
\caption{Time step convergence of guiding center second order Runge Kutta integrator. Panel.(a), Fig.(c) represent the relative energy error and panel.(b), Fig.(d) show relative canonical 
angular momentum error for DIII-D geometry core and cross-separatrix region, respectively.}
\label{fig:GCconvergence}
\end{figure*}
\begin{figure*}[htbp]
\centering
\includegraphics[width=\textwidth]{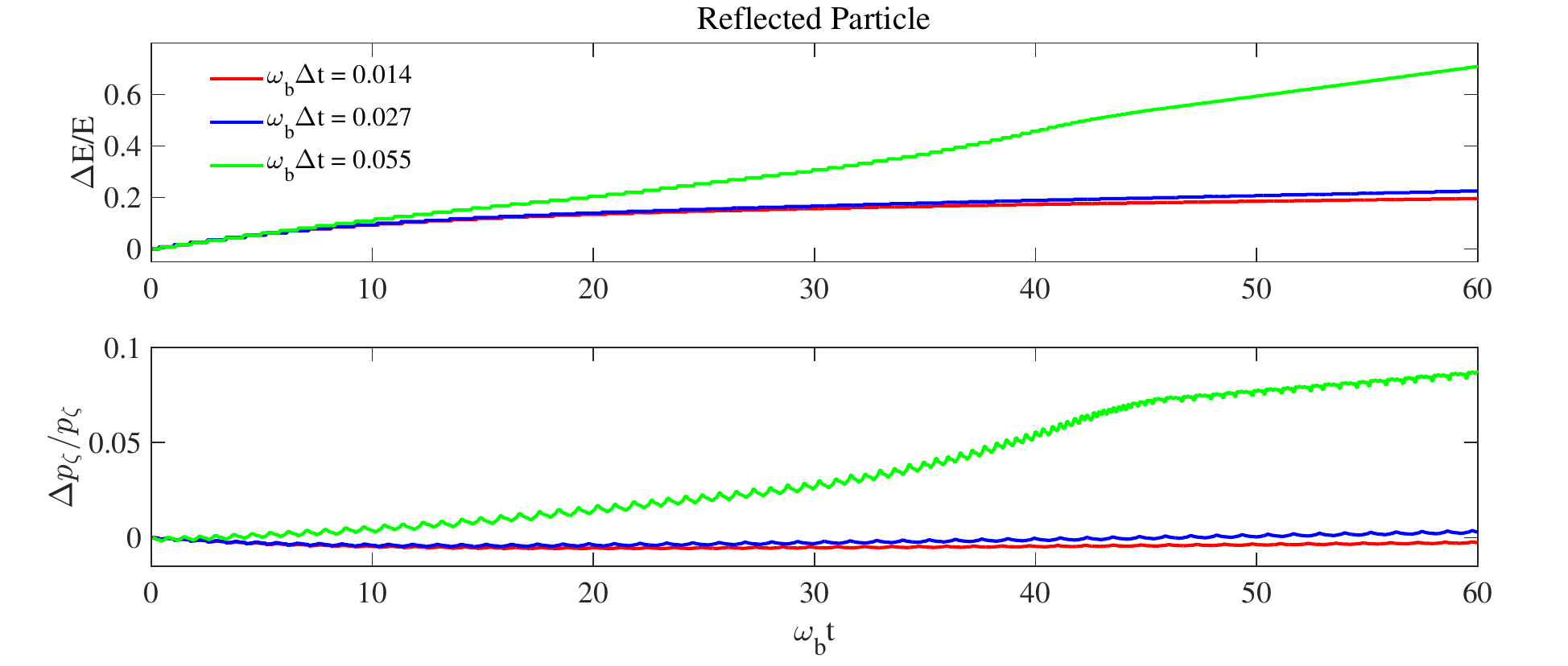}
\caption{Time step convergence of guiding center pusher for a boundary particle that is brought back into the simulation domain after leaving it. The figure on top shows the relative error in total energy while, the one below represents the relative error in the canonical angular momentum.}
\label{fig:Reflconvergence}
\end{figure*}

\subsection{Guiding center dynamics }
Guiding center particle dynamics is described by the five-dimensional phase space
\begin{equation}
\biggl[\frac{\partial}{\partial t}+\dot{\vec{X}}\cdot \nabla+
\dot v_\parallel\frac{\partial}{\partial v_\parallel}\biggr]f_{GC}=0,
\end{equation}
where $f_{GC}$ is the guiding center distribution function, $\vec{X}$ is the guiding center position and $v_\parallel$ is 
the parallel velocity. The evolution of the guiding center distribution function can be described by the following equations 
of guiding center motion\cite{brizard2007foundations}:  
\begin{eqnarray}
\dot{\vec{X}}=\frac{\vec B^*}{B_\parallel^*}v_\parallel +\vec{v}_{\Bbb E}+\vec{v}_c+\vec{v}_g,\nonumber\\
\dot v_\parallel=-\frac{1}{m}\frac{\vec B^*}{B_\parallel^*}\cdot(\mu\nabla B+q_c\nabla\phi),
\label{eq:GCvelocity}
\end{eqnarray}
where $\vec B^*=\vec B+B v_\parallel/\omega_c\nabla\times\hat b$, $\mu=mv_\perp^2/2B$, and $B_\parallel^*=\hat b\cdot \vec B^*$. The $\vec {\Bbb E}\times\vec B$ drift velocity $\vec v_{\Bbb E}$,
the grad-B drift velocity $\vec v_g$ and curvature drift velocity $\vec v_c$ are given by  
\begin{eqnarray}
\vec v_{\Bbb E}=\frac{c\vec B\times\nabla\phi}{BB_\parallel^*},\nonumber\\
\vec v_g=\frac{\mu}{m\omega_c}\frac{\vec B\times\nabla B}{B_\parallel^*},\nonumber\\
\vec v_c=\frac{B}{B_\parallel^*}\frac{v_\parallel^2}{\omega_c}\nabla\times\hat b.
\end{eqnarray}
In GC description following order is adopted:
\begin{eqnarray}
 \frac{\omega}{\omega_c}\sim \frac{k_\parallel}{k_\perp}\sim\frac{q_c\phi}{T_c}\sim \mathcal{O}(\epsilon),\nonumber
\end{eqnarray}
where $\omega$ is the frequency of the mode of interest, $k_\parallel$ and $k_\perp$ are the wave vectors in the parallel and perpendicular direction, respectively.Two constants of motion for guiding center dynamics are defined by:
\begin{itemize}
\item Total Energy
\begin{eqnarray}
\texttt{E}=(m/2)v_\parallel^2+\mu B,\nonumber
\end{eqnarray}
\end{itemize}
\begin{itemize}
\item Toroidal angular momentum 
\begin{eqnarray}
p_\zeta=mRv_\parallel(B_\zeta/B)+q_c\psi.\nonumber
\end{eqnarray}
\end{itemize}
For an axisymmetric system, Eq.~(\ref{eq:GCvelocity}) can be rewritten in cylindrical coordinate $(R,\zeta,Z)$ as follows:
\begin{eqnarray}
v^R=v_\parallel \frac{B_R}{B_\parallel^*}+\frac{c}{B_\parallel^*}\frac{B_\zeta}{B}\frac{\partial\phi}{\partial Z}
-\frac{B}{B_\parallel^*}\frac{v_\parallel^2}{\omega_c}\frac{\partial}{\partial Z}\biggl(\frac{B_\zeta}{B}\biggr)\nonumber\\
+\frac{\mu}{m\omega_c}\frac{B_\zeta}{B_\parallel^*}\frac{\partial B}{\partial Z}, 
\label{vr_gc}
\end{eqnarray}
\begin{eqnarray}
v^\zeta=v_\parallel \frac{B_\zeta}{B_\parallel^*}\frac{1}{R}
+\frac{c}{B_\parallel^*J}\biggl[\frac{B_Z}{B}\frac{\partial\phi}{\partial R}-\frac{B_R}{B}\frac{\partial\phi}{\partial Z}\biggr]\nonumber\\
+\frac{B}{B_\parallel^*}\frac{v_\parallel^2}{\omega_c}\frac{1}{J}\biggl[\frac{\partial}{\partial Z}\biggl(\frac{B_R}{B}\biggr)
-\frac{\partial}{\partial R}\biggl(\frac{B_Z}{B}\biggr)\biggr]\nonumber\\
\quad\quad\quad\quad\quad\quad-\frac{\mu}{m\omega_cJ}\biggl[\frac{B_R}{B_\parallel^*}\frac{\partial B}{\partial Z}
-\frac{B_Z}{B_\parallel^*}\frac{\partial B}{\partial R}\biggr],
\label{vzeta_gc}
\end{eqnarray}
\begin{eqnarray}
v^Z=v_\parallel \frac{B_Z}{B_\parallel^*}-\frac{c}{B_\parallel^*}\frac{B_\zeta}{B}\frac{\partial\phi}{\partial R}
+\frac{B}{B_\parallel^*}\frac{v_\parallel^2}{\omega_c}\frac{1}{J}\frac{\partial}{\partial R}\biggr(R\frac{B_\zeta}{B}\biggr)\nonumber\\
-\frac{\mu}{m\omega_c}\frac{B_\zeta}{B_\parallel^*}\frac{\partial B}{\partial R},
\label{vz_gc}
\end{eqnarray}
\begin{eqnarray}
\dot v_\parallel=-\frac{\mu}{m}\biggl(\frac{B_R}{B_\parallel^*}\frac{\partial B}{\partial R}+\frac{B_Z}{B_\parallel^*}\frac{\partial B}{\partial Z}\biggr)
-\frac{q_c}{m}\biggl(\frac{B_R}{B_\parallel^*}\frac{\partial \phi}{\partial R}+\frac{B_Z}{B_\parallel^*}\frac{\partial \phi}{\partial Z}\biggr)\nonumber\\
\quad-\frac{\mu v_\parallel}{m\omega_c}\frac{B}{B_\parallel^*}\frac{1}{J}\biggl[\frac{\partial}{\partial R}\biggl(R\frac{B_\zeta}{B}\biggr)\frac{\partial B}{\partial Z}
-R\frac{\partial}{\partial Z}\biggl(\frac{B_\zeta}{B}\biggr)\frac{\partial B}{\partial R}\biggr]\nonumber\\
\quad-\frac{v_\parallel q_c}{m\omega_c}\frac{B}{B_\parallel^*}\frac{1}{J}\biggl[\frac{\partial}{\partial R}\biggl(R\frac{B_\zeta}{B}\biggr)\frac{\partial \phi}{\partial Z}
-R\frac{\partial}{\partial Z}\biggl(\frac{B_\zeta}{B}\biggr)\frac{\partial \phi}{\partial R}\biggr].\nonumber\\
\label{vpara_gc}
\end{eqnarray}
For guiding center particle dynamics GTC normally uses second order Runge Kutta (RK) method 
(cf. Fig.2 of Ref. \cite{kuley2015verification}).

\begin{table}[htbp]
\centering
\caption{Initial conditions for FK integrator}
\begin{tabular}{ c c c }
\hline \hline
Parameter & Core  & Cross-Separatrix \\ \hline \hline
$R/R_0$   & 1.154  & 1.289 \\
$Z/R_0$   & 0.0 & 0.0    \\
 $\zeta$     & 1.570  & 1.570   \\
 $v^R/\omega_c R_0$   &3.371$\times 10^{-3}$  & 3.371$\times 10^{-3}$   \\
 $v^Z/\omega_c R_0$   &5.371$\times 10^{-3}$  & 5.371$\times 10^{-3}$    \\
 $v^\zeta/\omega_c $   &1.271$\times 10^{-3}$  & 1.271$\times 10^{-3}$  \\\hline 
   $R_0$ is the $R$ at magnetic axis
  \\\hline
 \hline
\end{tabular}
\label{Tbl:two_region_FK}
\end{table}

\begin{table}[htbp]
\centering
\caption{Initial conditions for GC integrator}
\begin{tabular}{ c c c }
\hline \hline
Parameter & Core  & Cross-Separatrix \\ \hline \hline 
$R/R_0$   & 1.147  & 1.281    \\
 $Z/R_0$   & 0.0 & 0.0    \\
 $\zeta$     & 1.570  & 1.570    \\
 $v_\parallel/\omega_c R_0$   &2.547$\times 10^{-3}$  & 3.017$\times 10^{-3}$   \\
 $\sqrt{\mu B_0}/\omega_c R_0$   & 4.545$\times 10^{-3}$  & 4.642$\times 10^{-3}$ \\\hline
  $B_0$ is the $B$ at magnetic axis
  \\\hline
  \hline
\end{tabular}
\label{Tbl:two_region_GC}
\end{table}

To test our integration schemes (FK and GC), we consider the collisionless single particle motion of the trapped one in the core and cross-separatrix regions using the initial conditions from Table \ref{Tbl:two_region_FK} and Table \ref{Tbl:two_region_GC} for DIII-D geometry, respectively. The projection of the FK and GC trapped particle orbit on R-Z plane in the core and cross-separatrix regions are shown in Fig.~\ref{fig:Orbit}. Both integrators correctly capture the trapped particle orbit and agree well with each other. In Fig.~\ref{torprec}, the temporal variation of the toroidal angle $\zeta$ of these orbits is shown. The bounce motion of the particles is implied by the oscillations of $\zeta$ and the gradual change in the mean level of these oscillations with time is an evidence of the toroidal precession of these orbits. The conservation properties of our integrators are tested with two exact constant of motion, viz., kinetic energy E and toroidal angular momentum $p_\zeta$.  

In numerical simulation the dynamical quantities accumulate error with time. Fig.~\ref{fig:FKconvergence} shows the relative variation of E and $p_\zeta$ for FK Boris integrator over a time of 15000 cyclotron period for different time step sizes in core and cross-separatrix regions. Boris algorithm maintains adequate accuracy with 20 time steps per cyclotron period $(\omega_c\Delta t=0.292)$. The relative energy error arises mostly due to floating point cutoff (floating precision). 
Whereas, the GC second order RK demonstrate that E and $p_\zeta$ can converge with 240 time steps per bounce period $(\omega_b\Delta t=0.027)$ in core and cross-separatrix regions, where $\omega_b$ is the bounce frequency of the particle [cf. Fig.(~\ref{fig:GCconvergence})]. From these convergence studies, it is found that the FK Boris integrator provides better energy convergence than the second order Runge Kutta GC integrator.  If there are high frequency electromagnetic perturbations, the condition of $\omega_{per}\Delta t< 1$ will set an upper bound for the time step size. In the present simulations, the particles orbit are studied without electric and perturbed magnetic field. However, previously Wei \textit{et al.}\cite{wei2015method} have demonstrated the effect of electric field on trapped particle orbit (Ware pinch) in the core region of tokamak using Boozer coordinates. 

As the particles move around, some of them happen to leave the domain of simulation. This, after a certain duration of time, would result in erroneous ensemble averages. To improve the accuracy of our simulations, we brought the escaping particles back into the simulation domain using the following procedure: We first found the flux at the position of the particle at a given instant of time $\psi_p(t)$, the simulation domain being the interval $(\psi_1,\psi_2)$, and the poloidal angle at that instant $\theta_p(t)$. If $\psi_p(t)<\psi_1$ or $\psi_p(t)>\psi_2$ i.e. if the particle lay outside the simulation domain, we re-initialized $\psi_p$ with its value at the earlier time step when the particle was within the simulation domain (i.e. $\psi_p(t)=\psi_p(t-\Delta t)$) and $\theta_p$ was re-initialized as $\theta_p(t)=2\pi - \theta_p(t-\Delta t)$. This technique preserves the nature of the motion of trapped particles at the domain boundaries. Knowing $\psi_p(t)$ and $\theta_p(t)$, we obtained the new coordinates $(R_p,Z_p)$ of the guiding center by finding expressions for $R$ and $Z$ in terms of $\psi$ and $\theta$. This was done by constructing 2D splines for $R=R(\psi,\theta)$ and $Z=Z(\psi,\theta)$ in the same way as mentioned in Section II.
\begin{eqnarray}
R(\psi,\theta)=R(1,i,j)+R(2,i,j)\Delta\psi +R(3,i,j)\Delta\psi^2 +\nonumber\\
R(4,i,j)\Delta\theta +R(5,i,j)\Delta\theta\Delta\psi +R(6,i,j)\Delta\theta\Delta\psi^2 +\nonumber\\
R(7,i,j)\Delta\theta^2 +R(8,i,j)\Delta\theta^2\Delta\psi +R(9,i,j)\Delta\theta^2\Delta\psi^2\nonumber\\
\label{eq:2DsplineR}
\end{eqnarray}
\begin{eqnarray}
Z(\psi,\theta)=Z(1,i,j)+Z(2,i,j)\Delta\psi +Z(3,i,j)\Delta\psi^2 +\nonumber\\
Z(4,i,j)\Delta\theta +Z(5,i,j)\Delta\theta\Delta\psi +Z(6,i,j)\Delta\theta\Delta\psi^2 +\nonumber\\
Z(7,i,j)\Delta\theta^2 +Z(8,i,j)\Delta\theta^2\Delta\psi +Z(9,i,j)\Delta\theta^2\Delta\psi^2\nonumber\\
\label{eq:2DsplineZ}
\end{eqnarray}
Here $\Delta\psi=\psi_{i+1}-\psi_i$ and $\Delta\theta=\theta_{j+1}-\theta_j$ where $i=1,2,...,N_1$ and $j=1,2,...,N_2$. $N_1$ and $N_2$ depend on the density of the spline grids that our accuracy of interpolation demands. The velocity of the escaping particle is re-initialized with its value at the previous time-step when the particle lay within the simulation domain. This was done to ensure the conservation of the total energy and the toroidal angular momentum.
In order to test the above technique, we applied it to the cross-separatrix particle in Fig. ~\ref{fig:Orbit}, using the separatrix as the boundary of our simulation domain such that the particle was brought back every time it crossed the separatrix. The relative error in the total energy and the canonical angular momentum for different time-steps is shown in Fig. ~\ref{fig:Reflconvergence}. We observe that the conservation of the invariants is better at the smaller time-step sizes ($\omega_b\Delta t=0.014, 0.027$) than that at a bigger time-step size ($\omega_b\Delta t=0.055$). However, the overall extent of conservation is not as good as observed earlier, in the absence of a domain boundary. This is because the boundary condition used here is not a perfect one. It however gives us good results in the self-consistent simulations of large number of particles over a long time intervals, as we will observe later in Section VI, where we verify zonal flows in the core region.

\begin{figure*}[htbp]
\centering
\includegraphics[width=\textwidth]{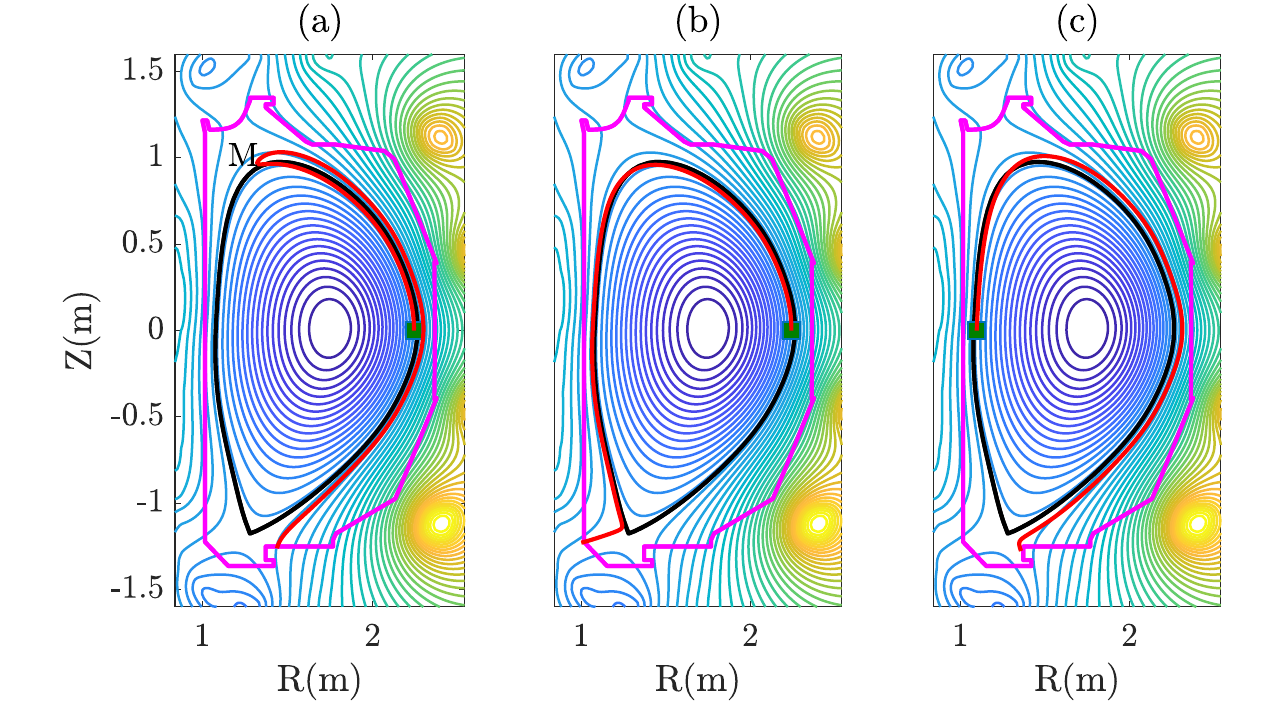}
\caption{Ion drift orbits (solid red line) of DIII-D with a single null divertor \cite{PhysRevLett.114.105002}. The ion toroidal grad-B drift is away from the X point ($B_\zeta>0$). Initial position of the test ion is represented by green square. (a) An ion starting from the outer midplane ($R_a>R_X$) is reflected at M, changes the sign of $v_\parallel$ and escapes near the X point to the outside divertor plate [case-I]. (b) An ion starting from the outer midplane escapes without mirror reflection to the inside divertor plate [case-II]. (c) An ion starting from the inner midplane $(R_a<R_X)$ escapes near the X point to the outside divertor plate [case-III]. Last close surface and limiter points are represented by black and magenta line, respectively.}
\label{fig:orbitloss}
\end{figure*}

 
 \section{Ion orbit loss near plasma edge}
 \label{ion_orbit_loss}
In a tokamak plasma, ions have drift motions due to the gradient-B and curvature drift. As a result, ion orbits are shifted from a magnetic surface. Due to this shift of ion orbit from magnetic surface, the hot ions that exist close to the separatrix, can pass near the X point region. In this region, the poloidal magnetic field is very weak and the ions have a very small poloidal displacement in time. These ions experience vertical curvature and grad-B drifts and move towards the divertor, resulting in ion orbit loss \cite{pan2014co,miyamoto1996direct}. In an axisymmetric configuration, the drift orbit of ion is obtained by solving the following three equations as described in Sec 4.3. These are the conservation of kinetic energy, magnetic moment and canonical angular momentum.
\begin{equation}
\frac{1}{2}m(v_\perp^2+v_\parallel^2)+q_c\phi=\texttt{constant}=\frac{1}{2}m(v_{\perp a}^2+v_{\parallel a}^2)+q_c\phi_a,
\end{equation}
\begin{equation}
\frac{mv_{\perp}^2}{2B}=\texttt{constant}=\frac{mv_{\perp a}^2}{2B_a},
\end{equation}
\begin{equation}
mRv_\parallel (B_\zeta/B)+q_c\psi=\texttt{constant}=mR_av_{\parallel a} (B_{\zeta a}/B_a)+q_c\psi_a.
\label{eq:momentumconservation}
\end{equation}

\begin{table*}[htbp]
\centering
\caption{Parameters for ion orbit loss region}
\begin{tabular}{ c c c }
\hline \hline
Parameter & Case-I & Case-II  \\
\hline \hline
$(R_a/R_0, Z_a/R_0)$   & (1.2748, 0.0)  & (1.2748, 0.0)    \\
 $(R_{in}/R_0, Z_{in}/R_0)$   & (0.6134, 0.0) & (0.6134, 0.0)    \\
 $(R_{XD}/R_0, Z_{XD}/R_0)$     & (0.8442, -0.6680)  & (0.6253, -0.6680)    \\
 $(\psi_{XD}/B_0R_0^2, \psi_a/B_0R_0^2)$   & (5.253$\times 10^{-2}$, 4.688$\times 10^{-2}$)  & (5.138$\times 10^{-2}$, 4.688$\times 10^{-2}$) \\
 \hline 
\end{tabular}
\label{Tbl:two_region_ion}
\end{table*}

The subscript `a' means the value at the starting point of the ions. The conservation of above quantities with $B_\zeta\simeq B$ gives the drift orbit surface as follows \cite{miyamoto1996direct}: 
\begin{equation}
\frac{q_c\psi}{mv_a}+R\frac{v_\parallel}{v_a}=\frac{q_c\psi_a}{mv_a}+R_a\frac{v_{\parallel a}}{v_a},
\label{eq:driftsurface}
\end{equation}
\begin{equation}
\frac{v_\parallel}{v_a}=\pm\biggl[1-(1-\xi_0^2)\frac{R_a}{R}+\frac{q_c}{mv_a^2/2}(\phi_a-\phi)\biggr]^{1/2},
\label{eq:pitchangle}
\end{equation}
where $\xi_0=(v_{\parallel a}/v_a)$ is the initial pitch angle, and $\xi_0>0$ denotes the ion moving in the co-current direction.  In the above equation, we assume that $B_\zeta>0$, and use the approximation $B_\zeta\propto(1/R)$. A minus sign must be used instead of a plus sign in Eq.~(\ref{eq:momentumconservation}) when $B_\zeta<0$. We also consider that the poloidal magnetic field is directed clockwise and the ion toroidal drift $(\vec{v}_g)$ is approximately vertically up and away from the X point. The coordinates of the X point are $(R_X,Z_X)$ and the poloidal magnetic flux function of X point satisfies $(\partial\psi/\partial R)_X=0$ and $(\partial\psi/\partial Z)_X=0$.  When the initial position $R_a$ of a test ion is at the outer midplane, that is $R_a>R_X$ there are following two cases for orbit ion loss [cf. Fig.~\ref{fig:orbitloss}(a)-(b)] \cite{miyamoto1996direct}:

\begin{figure*}[htbp]
\centering
\includegraphics[width=\textwidth]{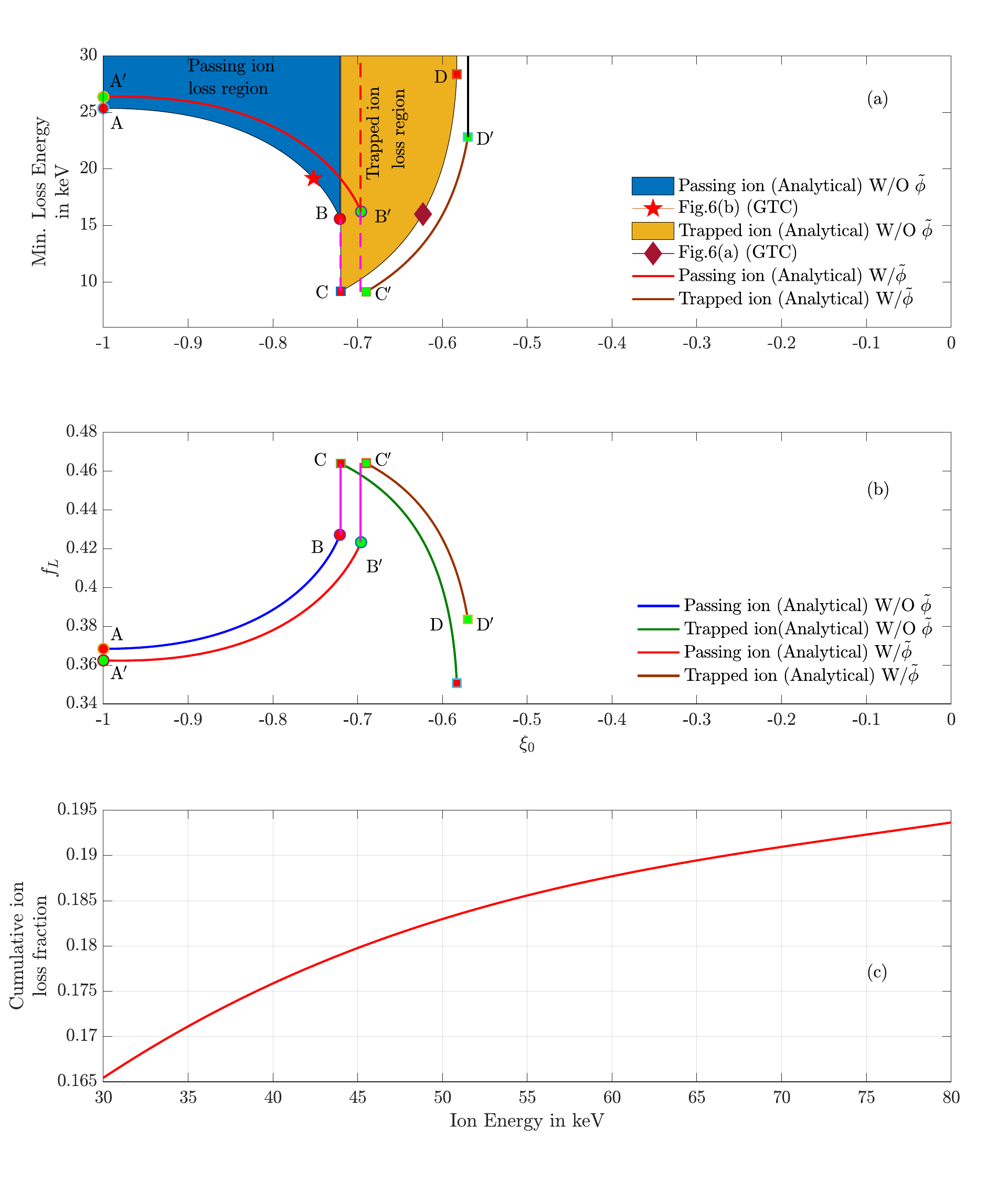}
\caption{Ion orbit loss region for DIII-D \cite{PhysRevLett.114.105002} in the initial velocity space for the same parameters of Fig.8. (a) The minimum energy for which the orbits of ions launched from $R_a>R_X$ will be lost. (b) Pitch angle dependent ion orbit loss fraction for $T_i$=40 keV. Parts AB (A$'$B$'$), BC (B$'$C$'$) and CD (C$'$D$'$) of the curve ABCD (A$'$B$'$C$'$D$'$) correspond to the conditions Case-II, Case-I(ii) and Case-I(iii) without $\tilde{\phi}$ (with $\tilde{\phi}$), respectively. (c) Cumulative ion orbit loss fraction as a function of ion energy ($T_i$).}
\label{fig:orbitlosscomparison}
\end{figure*}

When $\xi_0>0$ and $R_a>R_X$, the relative position of magnetic surface and drift orbit surface [cf. Eq.~(\ref{eq:driftsurface})], clearly indicates that direct ion orbit loss is not possible. However, when the initial position of the test ion is selected as $R_a<R_X$ the following Fig.~\ref{fig:orbitloss}(c) is possible for ion orbit loss. 

 In the above GTC-X simulations (cf. Fig.~\ref{fig:orbitloss}) we did not include the radial electric field on ion orbit loss. However, one must mention here that in a tokamak scenario the radial electric field shifts the velocity space boundaries separating trapped and passing orbits. Also the ion orbit loss contributes to the generation of radial electric field, which affects the ion orbit loss. Therefore, the ion orbit loss and the radial electric field should be considered self-consistently for more accurate calculation. Similarly, it is important to study the effect of ion orbit loss and X point loss for understanding the plasma transport \cite{stacey2011effect}, particle distribution \cite{stacey2013effect}, plasma rotation \cite{pan2014co}, etc. 
Finally, the ion orbit loss region in our calculation is based on the assumptions that the ions are collisionless and the loss region is not filled by the collisions or turbulence. Simulation related to these effects will be reported in a future work. To see how the electric field affects the velocity space boundaries, we have calculated the minimum velocity of the ion and particle loss fractions for DIII-D in the following paragraph. 

In order to find the minimum initial ion speed $v_0$ that is required for the ion to reach the final location (near X point) we combine Eq.~(\ref{eq:driftsurface}) and Eq.~(\ref{eq:pitchangle}). The minimum loss speed of ion is 
\begin{widetext}
\begin{eqnarray}
v_0=\bigg[R_a^2\xi_0^2-R_{XD}^2\biggl\{1-(1-\xi_0^2)\frac{R_a}{R_{XD}}\biggr\}\biggr]^{-1}\times
\label{eq:minvelocity}
\\
-\Psi R_a\xi_0\pm\sqrt{\Psi^2R_a^2\xi_0^2-\biggl[\Psi^2-R_{XD}^2\tilde{\phi}\biggr]\bigg[R_a^2\xi_0^2-R_{XD}^2\biggl\{1-(1-\xi_0^2)\frac{R_a}{R_{XD}}\biggr\}\biggr]}\nonumber
\end{eqnarray}
\end{widetext}
where $\Psi=(q_c/m)(\psi_a-\psi_{XD})$ and $\tilde \phi=q_c(\phi_a-\phi_{XD})/(m\omega_c^2R_0^2/2)$. For passing ion orbit loss the sign of the second term of the numerator in Eq.~(\ref{eq:minvelocity}) is positive and for the trapped ion orbit loss it is negative, respectively. We solve Eq.~(\ref{eq:minvelocity}) to find the ion orbit loss region for DIII-D model parameters as described in Table \ref{Tbl:two_region_ion}. Here, we examine the loss of ion for which $-1\le\xi_0\le 0$. The loss region in the initial velocity space of an ion starting from outer midplane point $R_a$ is described by the solid curves in Fig.~\ref{fig:orbitlosscomparison}(a). Case-I and Case-II represent boundary curves CD (C$'$D$'$) and AB (A$'$B$'$) of the loss region without $\tilde{\phi}$ (with $\tilde\phi$), respectively.  Boundary curves BC (B$'$C$'$) i.e, the condition for mirror reflection of the test ion without $\tilde{\phi}$ (with $\tilde\phi$) is given by Case-I(ii),where $\tilde{\phi}>0$ and $\tilde{\phi}<0$ correspond to the outward and inward radial electric field, respectively.  The value of minimum energy $(mv_0^2/2)$ is obtained from Eq.~(\ref{eq:minvelocity}) [cf. Fig.~\ref{fig:orbitlosscomparison}(a)] without $\tilde\phi$ demonstrate good agreements with the simulation results of Fig.~\ref{fig:orbitloss}(a) and Fig.~\ref{fig:orbitloss}(b).

After the minimum loss energy is determined, particle loss fraction due to direct ion orbit loss is calculated. The ion loss rate is determined by the rate of supply of ions to the loss regions. The corresponding cumulative particle loss fraction of ions, which follow distribution function $g$ in velocity space is defined as \cite{miyamoto1996direct}:
\begin{equation}
f_L=\int_{-1}^1\frac{\int_{v_0{(\xi_0)}}^\infty v^2g(v)dv}{\int_{-\infty}^\infty v^2g(v)dv}d\xi_0
\end{equation} 
For a Maxwellian distribution function the above equation turns out to be, 
 $f_L=\int_{-1}^1 \Gamma(3/2,\epsilon_{min}(\xi_0))d\xi_0/2\Gamma(3/2)$, where $\epsilon_{min}(\xi_0)=mv_0^2(\xi_0)/2T_i$ is the energy corresponding to the minimum velocity for which ion orbit loss is possible [cf. Eq.~(\ref{eq:minvelocity})], $T_i$ is the ion temperature, and $\Gamma(3/2,\epsilon_{min}(\xi_0))$ is the upper incomplete gamma function of order 3/2. The $\xi_0$ dependent cumulative particle loss fraction $f_L$ starting from $R_a>R_X$ without $\tilde{\phi}$ (ABCD) and with $\tilde\phi$ (A$^\prime$B$^\prime$C$^\prime$D$^\prime$) are shown in Fig.~\ref{fig:orbitlosscomparison}(b). These results clearly demonstrate that radial electric field shifts the velocity space boundaries of loss regions for trapped and passing particles. Fig.~\ref{fig:orbitlosscomparison}(c) represents the cumulative ion loss fraction for different ion energy ($T_i$). Pan et al., have carried out detailed studies of minimum loss energy and cumulative loss fraction for the ions starting from different poloidal and radial positions \cite{pan2014co}. However, for verification and application of our new capability we have carried out the above calculation for an ion starting from the outer mid-plane only.
 
 \section{Verification of Zonal flows in the core region} 
 \label{zonal_section}
 
 In this section, we demonstrate the collisionless damping of zonal flows and verify the Rosenbluth-Hinton theory \cite{Rosenbluth} by observing the variation of the zonal electric field $E(\psi)$, at a given point in the core region, with respect to time. Zonal flows are low frequency electrostatic modes that are spontaneously generated by turbulence and, in turn, play an important role in regulating the turbulence. They are important in the study of transport processes in the core region. In order to simulate these zonal flows, we solve the flux-surface averaged gyrokinetic Poisson's equation \cite{xiao2015gyrokinetic} at every time step.
 \begin{equation}
 \langle\nabla_\perp^2\phi\rangle=\biggl<\biggl(\frac{T_i}{Z_i^2n_i}\nabla_\perp^2-\frac{T_i}{\rho_i^2Z_i^2n_i}\biggr)\frac{1}{e}(Z_in_i-n_e)\biggr>
\label{eq:gyrokineticPoisson}
 \end{equation}
 
 \begin{figure*}[htbp]
 \begin{center}
 \includegraphics[width=\textwidth]{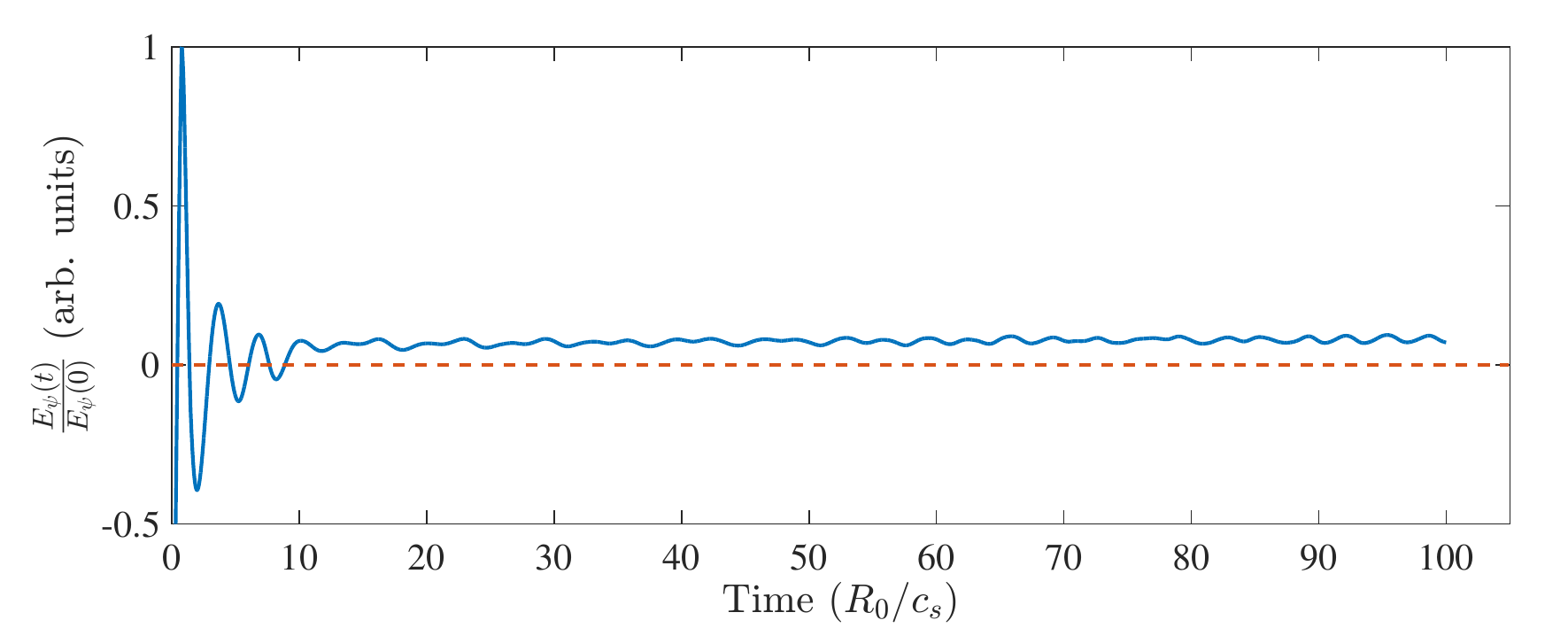}
 \caption{Time series plot of the normalized zonal electric field on a given flux surface. Here, $R_0$ is the major radius of the DIII-D tokamak and $c_s$ is the ion sound speed.}
 \label{zonalplot}
 \end{center}
 \end{figure*}
 
 Here $\langle...\rangle$ implies flux-surface average. $T_i, n_i$ and $Z_i$ are the temperature, density and atomic number of the ion species respectively. $e$ is the unit electronic charge and $n_e$ is the electron density. $\rho_i$ is the ion gyroradius. Since we are dealing with zonal quantities, we are only concerned with the component of the Laplcian perpendicular to the flux  surfaces. As shown in Xiao \textit{et al.}\cite{xiao2015gyrokinetic}, the flux-surface averaged Laplacian can be written in the flux coordinates $(\psi,\theta,\zeta)$ as
\begin{equation}
\langle\nabla_\perp^2\phi\rangle=\frac{1}{J_0(\psi)}\biggl[\frac{\partial}{\partial\psi}\biggl(J_0(\psi)\langle g^{\psi\psi}\rangle\frac{\partial\langle\phi\rangle}{\partial\psi}\biggr)\biggr]
\label{eq:fluxcoordLapl}
\end{equation}
where $J_0(\psi)$ is the flux-surface averaged Jacobian. In writing Eq.(\ref{eq:fluxcoordLapl}) we have neglected the higher order toroidal coupling terms by assuming a small inverse-aspect-ratio. In our simulations we load a uniform ion temperature and density profile. As a result, $T_i$ and $n_i$ are constant all throughout the simulation domain. Integrating Eq.(\ref{eq:gyrokineticPoisson}) once with respect to $\psi$, we obtain the zonal electric field as follows:
\begin{eqnarray}
E(\psi)=\frac{\partial\langle\phi\rangle}{\partial\psi}=-\frac{T_i}{n_iZ_i^2e^2}\frac{\partial\langle\rho_c\rangle}{\partial\psi}\nonumber\\
-\frac{1}{\langle g^{\psi\psi}\rangle J_0(\psi)}\int d\psi\frac{T_i}{n_iZ_i^2e^2}\biggl<\frac{1}{\rho_i^2}\biggr>\langle\rho_c\rangle J_0(\psi)
\label{eq:zonalField}
\end{eqnarray}

Here $\langle\rho_c\rangle = e(Z_i\langle\overline{n}_i\rangle -\langle n_e\rangle )$ is the flux-averaged total charge density. $\overline{n}_i$ is the ion guiding center density. Solving Eq.(\ref{eq:zonalField}) at every time step, we have incorporated the field information in our GC pusher which updates the particle position accordingly. However, Eq.(\ref{eq:zonalField}) gives the zonal electric as a function of $\psi$ while the guiding center equations of motion in cylindrical coordinates (Eqs. \ref{vr_gc}-\ref{vpara_gc}) require the $R$ and $Z$ components of the zonal field. Hence we simply do a coordinate transformation to obtain $E_R$ and $E_Z$ i.e. $E_R=(\partial\psi/\partial R)E_\psi$ and $E_Z=(\partial\psi/\partial Z)E_\psi$, where $E_\psi=E(\psi)$.

In our simulation of zonal flows, we started off by loading the particles (here, thermal ions) in a certain domain within the core region such that, they are uniformly distributed between any two consecutive flux surfaces. For this, we had to calculate the volume associated with each flux surface. We began by taking an approximation of the volume associated with each grid cell and then added them up over the respective flux surfaces to get the flux surface volumes. After that we loaded the particles accordingly and calculated the number of particles lying within a certain neighborhood of each flux surface (i.e. their weights associated with a given flux surface), thereby obtaining the flux surface volumes in terms of the number of particles. The newly calculated flux surface volumes would also serve as the flux-averaged Jacobian $J_0(\psi)$ at each flux surface. As the system evolves with time, the particles leaving the simulation domain are brought back using the reflective boundary conditions discussed earlier, in order to maintain quasi-neutrality and energy conservation. Full-f ions and adiabatic electrons are used in the present study of zonal flows. The initial velocities of the particles followed a Maxwellian distribution. The simulation domain extended from $\psi_1=0.4\psi_{sep}$ to $\psi_2=0.9\psi_{sep}$ and the domain-width along the outer midplane was $\Delta\approx68.02\rho_i$.

The zonal flows were initiated by driving the system for a certain duration of time with a low magnitude electric field having a sinusoidal dependence on $\psi$. After switching off the external perturbation (i.e. the driving field) the system was allowed to evolve self-consistently with time. Fig. \ref{zonalplot} shows the time evolution of the zonal field $E_\psi$ after removing the external perturbation. Finally $E_\psi$ settles down to a residual value, in the absence of collisions. This non-zero value is the Rosenbluth-Hinton residual level and is of great importance in determining the level of turbulence in the plasma \cite{Rosenbluth}. 

\section{Summary and Conclusion}
\label{disc}
As a first step in developing the self-consistent global simulation model to couple the SOL and the core regions in tokamaks by incorporating the sepratrix, we have developed the particle dynamics for FK and GC particles using cylindrical coordinates. To get the maximum numerical efficiency, field aligned mesh in the cylindrical coordinates is developed for the whole device from magnetic axis to the material wall (plasma facing components). These field-aligned particle-grid interpolations using an axisymmetric mesh in cylindrical coordinates helps to avoid the difficulty associated with the X point. Finally, we have extended this particle orbit simulation techniques to study the ion orbit loss near the X-point of DIII-D single null divertor. 
Based on three constants of motion, the minimum loss speed and ion orbit loss fraction at the edge of the tokamak are calculated.  
However, these particles need proper boundary conditions near the divertor wall. Finally, we also verified the physics of zonal flows using the equlibrium of the  DIII-D tokomak. Presently GTC-X does not have the capability of handling the plasma particle loss to the material wall (plasma facing components), recycling the neutral particles, and calculating the Monte-Carlo neutral particle transport with a charge exchange and ionization interaction with the plasma. 

\section{Acknowledgements}
A.K. would like to thank Dr. R. Nazikian and Ms. D. Sharma for providing EFIT and IPREQ data. This work is supported by Board of Research in Nuclear Sciences (BRNS Sanctioned $\#$ 39/14/05/2018-BRNS), Science and Engineering Research Board EMEQ program (SERB Sanctioned $\#$ EEQ/2017/000164), US Department of Energy (DOE)
SciDAC ISEP Program, China National Magnetic Confinement Fusion Energy Research Program,
(Grant No. 2013GB111000) and IISc startup grant. A.S. thanks the Indian National Science
Academy (INSA) for their support under the INSA Senior Scientist Fellowship scheme. All the simulations were carried out on Cray XC40 - SahasraT at Supercomputing Education and Research Centre (SERC), IISc.


\begin{thebibliography}{99}
 \bibitem{loarte2007power} A. Loarte, B. Lipschultz, A. S. Kukushkin, G. F. Matthews, P. C. Stangeby, N. Asakura, G. F. Counsell, G. Federici, A. Kallenbach, K. Krieger, and others, \href{https://doi.org/10.1088/0029-5515/47/6/S04}{Nucl. Fusion} \textbf{47}, S203 (2007).
 \bibitem{cesario2010current}R. Cesario, L. Amicucci, A. Cardinali, C. Castaldo, M. Marinucci, L. Panaccione, F. Santini, O. Tudisco, M. L. Apicella, G. Calabro, C. Cianfarani, D. Frigione, A. Galli, G. Mazzitelli, C. Mazzotta, V. Pericoli, G. Schettini, A. A. Tuccillo, and the FTU Team, \href{https://doi.org/10.1038/ncomms1052}{Nat. Commun.} \textbf{1}, 55 (2010).
\bibitem{wising1996simulation} F. Wising, D. A. Knoll, S. I. Krasheninnikov, T. D. Rognlien, and D. J. Sigmar, \href{https://doi.org/10.1002/ctpp.2150360238}{Contributions to Plasma Phys.} \textbf{36}, 309 (1996).
\bibitem{rozhansky2009new} V. Rozhansky, E. Kaveeva, P. Molchanov, I. Veselova, S. Voskoboynikov, D. Coster, G. Counsell, A. Kirk,  S. Lisgo, and MAST Team and others, \href{https://doi.org/10.1088/0029-5515/49/2/02500}{Nucl. Fusion} \textbf{49}, 025007 (2009). 
\bibitem{braginskii1965review} S. I. Braginskii, Review of Plasma Physics,  \textbf{1},  205 (1965).
\bibitem{chankin2007discrepancy} A. V. Chankin, D. P. Coster, N.  Asakura, X. Bonnin, G. D. Conway, G. Corrigan, S. K. Erents, W. Fundamenski, J. Horacek, A. Kallenbach, and others, \href{https://doi.org/10.1088/0029-5515/47/5/013}{Nucl. Fusion} \textbf{47}, 479 (2007).
\bibitem{erents2004comparison}  S. K. Erents, R. A. Pitts, W. Fundamenski, J. P. Gunn, and G. F. Matthews, \href{https://doi.org/10.1088/0741-3335/46/11/006}{Plasma Phys. Control. Fusion}, \textbf{46}, 1757 (2004). 
 \bibitem{wischmeier2011assessment} M. Wischmeier, M. Groth, S. Wiesen, S. Potzel, L. Aho-Mantila, D. P. Coster, R. Dux, C. Fuchs, A. Kallenbach, H. W. M{\"u}ller, and others, \href{https://doi.org/10.1016/j.jnucmat.2011.02.020}{Journal of Nucl. Materials} \textbf{415}, S523 (2011).
 \bibitem{pan2014co} C. Pan, S. Wang, and J. Ou, \href{doi:10.1088/0029-5515/54/10/103003}{Nucl. Fusion} \textbf{54}, 103003 (2014).
 \bibitem{chang2002x}C. S. Chang, S. Kue, and H. Weitzner, \href{https://doi.org/10.1063/1.1490348}{Phys. Plasmas} \textbf{9}, 3884 (2002).
 \bibitem{omotani2013non} J. T. Omotani and B. D. Dudson, \href{https://doi.org/10.1088/0741-3335/55/5/055009}{Plasma Phys. Control. Fusion} \textbf{55}, 055009 (2013).
 \bibitem{wesson1995effect} J. A. Wesson, \href{https://doi.org/10.1088/0741-3335/37/12/008}{Plasma Phys. Control. Fusion} \textbf{37}, 1459 (1995).
 \bibitem{kuley2009stabilization} A. Kuley and V. K. Tripathi, \href{https://doi.org/10.1063/1.3080744}{Phys. Plasmas} \textbf{16}, 032504 (2009).
 \bibitem{kuley2010parametric} A. Kuley and V. K. Tripathi, \href{https://doi.org/10.1063/1.3442745}{Phys. Plasmas} \textbf{17}, 062507 (2010).
 \bibitem{kuley2010lower} A. Kuley, C. S. Liu, and V. K. Tripathi, \href{https://doi.org/10.1063/1.3454692}{Phys. Plasmas} \textbf{17}, 072506 (2010).
 \bibitem{chang2009compressed} C. S. Chang, S. Ku, P. H. Diamond, Z. Lin, S. Parker, T. S. Hahm, and N. Samatova, \href{https://doi.org/10.1063/1.3099329}{Phys. Plasmas} \textbf{16}, 056108 (2009).
 \bibitem{lin1998turbulent}Z. Lin, T. S. Hahm, W. W. Lee, W. M. Tang, and R. B. White, \href{https://doi.org/10.1126/science.281.5384.1835}{Science} \textbf{281}, 1835 (1998).
 \bibitem{lin1997neoclassical} Z. Lin, W. Tang, and W. Lee, \href{https://doi.org/10.1103/PhysRevLett.78.456}{Phys. Rev. Lett.} \textbf{78}, 456 (1997).
 \bibitem{dong2016effects} G. Dong and Z. Lin, \href{https://doi.org/10.1088/1741-4326/57/3/036009}{Nucl. Fusion} \textbf{57}, 036009 (2016).
 \bibitem{xie2017new} H. Xie, Y. Xiao, and Z. Lin, \href{https://doi.org/10.1103/PhysRevLett.118.095001}{Phys. Rev. Lett.} \textbf{118}, 095001 (2017).
 \bibitem{fulton2016gyrokinetic} S. Taimourzadeh, L. Shi, Z. Lin, R. Nazikian, I. Holod, D. Spong, \href{https://doi.org/10.1088/1741-4326/aafe3a}{Nucl. Fusion} \textbf{59}, 046005 (2019). 
 \bibitem{xiao2009turbulent} Y. Xiao and Z. Lin, \href{https://doi.org/10.1103/PhysRevLett.103.085004}{Phys. Rev. Lett.} \textbf{103}, 085004 (2009).
 \bibitem{holod2016effect} I. Holod, Z. Lin, S. Taimourzadeh, R. Nazikian, D. Spong, and A. Wingen, \href{https://doi.org/10.1088/0029-5515/57/1/016005}{Nucl. Fusion} \textbf{57}, 016005 (2017).
 \bibitem{xiao2015gyrokinetic} Y. Xiao, I. Holod, Z. X. Wang, Z. Lin, T. G. Zhang, \href{https://doi.org/10.1063/1.4908275}{Phys. Plasmas} \textbf{22}, 022516 (2015).
 \bibitem{zhang2012nonlinear} H. S. Zhang, Z. Lin, and I. Holod, \href{https://doi.org/10.1103/PhysRevLett.109.025001}{Phys. Rev. Lett.} \textbf{109}, 025001 (2012).
 \bibitem{spong2012verification} S. Taimourzadeh, E. M. Bass, Y. Chen, C. Collins, N. N. Gorelenkov, A. Könies, Z. X. Lu 8, D. A. Spong, Y. Todo, M. E. Austin, J. Bao, A. Biancalani, M. Borchardt, A. Bottino, W. W. Heidbrink, R. Kleiber, Z. Lin, A. Mishchenko, L. Shi, J. Varela, R. E. Waltz, G. Yu, W. L. Zhang, Y. Zhu, \href{https://doi.org/10.1088/1741-4326/ab0c38}{Nucl. Fusion} \textbf{59}, 066006 (2019). 
 \bibitem{wang2013radial} Z. X. Wang, Z. Lin, I. Holod, W. W. Heidbrink, B. Tobias, M. V. Zeeland, and M. E. Austin, \href{https://doi.org/10.1103/PhysRevLett.111.145003}{Phys. Rev. Lett.} \textbf{111}, 145003 (2013). 
 \bibitem{mcclenaghan2014verification}J. McClenaghan, Z. Lin, I. Holod, W. Deng, and Z. Wang, \href{https://doi.org/10.1063/1.4905073}{Phys. Plasmas} \textbf{21}, 122519 (2014).
 \bibitem{liu2014verification}D. Liu, W. Zhang, J. McClenaghan, J. Wang, Z. Lin, \href{https://doi.org/10.1063/1.4905074}{Phys. Plasmas} \textbf{21}, 122520 (2014).
 \bibitem{bao2017conservative} J. Bao, D. Liu, and Z. Lin, \href{https://doi.org/10.1063/1.4995455}{Phys. Plasmas} \textbf{24}, 102516 (2017).
 \bibitem{kuley2013verification} A. Kuley, Z. X. Wang, Z. Lin, and F. Wessel, \href{https://doi.org/10.1063/1.4826507}{Phys. Plasmas} \textbf{20}, 102515 (2013).
 \bibitem{bao2014particle} J. Bao, Z. Lin, A. Kuley, and Z. X. Lu, \href{https://orcid.org/0000-0001-7946-8682}{Plasma Phys. and Control. Fusion} \textbf{56}, 095020 (2014).
 \bibitem{kuley2015verification} A. Kuley, Z. Lin, J. Bao, X. S. Wei, Y. Xiao, W. Zhang, G. Y. Sun, and N. J. Fisch, \href{https://doi.org/10.1063/1.4934606}{Phys. Plasmas} \textbf{22}, 102515 (2015).
 \bibitem{kuley2015nonlinear} A. Kuley, J. Bao, Z. Lin, X. S. Wei, Y. Xiao, Proceedings of the 21st Topical Conference on Radiofrequency Power in Plasmas, \href{http://dx.doi.org/10.1063/1.4936506}{AIP Conference Proceedings} \textbf{1689}, 060008 (2015).
 \bibitem{bao2015global}J. Bao, Z. Lin, A. Kuley, Proceedings of the 21st Topical Conference on Radiofrequency Power in Plasmas, \href{http://dx.doi.org/10.1063/1.4936531}{AIP Conference Proceedings} \textbf{1689}, 080008 (2015).
 \bibitem{bao2016electromagnetic} J. Bao, Z. Lin, A. Kuley, and Z. X. Wang, \href{https://doi.org/10.1088/0029-5515/56/6/066007}{Nucl. Fusion} \textbf{56}, 066007 (2016).
 \bibitem{bao2016nonlinear} J. Bao, Z. Lin, A. Kuley, and Z. X. Wang, \href{https://doi.org/10.1063/1.4952773}{Phys. Plasmas} \textbf{23}, 062501 (2016).
 \bibitem{fulton2016bgyrokinetic} D. P. Fulton, C. K. Lau, I. Holod, Z. Lin, and S. Dettrick, \href{https://doi.org/10.1063/1.4930289}{Phys. Plasmas} \textbf{23}, 012509 (2016).
\bibitem{lao1990equilibrium} L. L. Lao, J. R. Ferron, R. J. Groebner, W. Howl, H. St John, E. J. Strait, and T. S. Taylor, \href{https://doi.org/10.1088/0029-5515/30/6/006}{Nucl. Fusion} \textbf{30}, 1035 (1990). 
 \bibitem{ren2011high} Q. Ren, M. Chu, L. Lao, and R. Srinivasan, \href{https://doi.org/10.1088/0741-3335/53/9/095009}{Plasma Phys. and Control. Fusion} \textbf{53}, 095009 (2011). 
 \bibitem{Srinivasan} Report on ``Tokamak Equilibrium Code - IPREQ" by R. Srinivasan and S. P. Deshpande.
 \bibitem{Bao19} J. Bao, C. K. Lau, Z. Lin, H. Y. Wang, D. P. Fulton, S. Dettrick, T. Tajima, \href{https://doi.org/10.1063/1.5087079}{Phys. Plasmas} \textbf{26}, 042506 (2019).
 \bibitem{kim2017happens} K. Kim, C. S. Chang, J. Seo, S. Ku, and W. Choe, \href{https://doi.org/10.1063/1.4974777}{Phys. Plasmas} \textbf{24}, 012306 (2017). 
 \bibitem{hariri2013flux} F. Hariri and M. Ottaviani, \href{http://dx.doi.org/10.1016/j.cpc.2013.06.005}{Computer Phys. Communications} \textbf{184}, 2419 (2013).
 \bibitem{Sturdevant2017} B. J. Sturdevant, Y. Parker, and S. E. Parker, \href{https://doi.org/10.1063/1.4999945}{Phys. Plasmas} \textbf{24}, 081207 (2017).
 \bibitem{miyamoto1996direct} K. Miyamoto, \href{https://doi.org/10.1088/0029-5515/36/7/I09}{Nucl. Fusion} \textbf{36}, 927 (1996).
 \bibitem{stacey2011effect} W. M. Stacey, \href{https://doi.org/10.1063/1.3640506}{Phys. Plasmas} \textbf{18}, 102504 (2011).
  \bibitem{stacey2013effect} W. M. Stacey,  \href{https://doi.org/10.1088/0029-5515/53/6/063011}{Nucl. Fusion} \textbf{53}, 063011 (2013).
 \bibitem{Rosenbluth} M. N. Rosenbluth, F. L. Hinton, \href{https://doi.org/10.1103/PhysRevLett.80.724}{Phys. Rev. Lett.} \textbf{80}, 724 (1998).
 \bibitem{PhysRevLett.114.105002} R. Nazikian, C. Paz-Soldan, J. D. Callen, J. S. deGrassie, D. Eldon, T. E. Evans, N. M. Ferraro, B. A. Grierson, R. J. Groebner, S. R. Haskey, C. C. Hegna, J. D. King, N. C. Logan, G. R. McKee, R. A. Moyer, M. Okabayashi, D. M. Orlov, T. H. Osborne, J-K. Park, T. L.  Rhodes, M. W. Shafer, P. B. Snyder, W. M. Solomon, E. J. Strait, and M. R. Wade, \href{https://doi.org/10.1103/PhysRevLett.114.105002}{Phys. Rev. Lett.} \textbf{114}, 105002 (2015).
 \bibitem{wiki.org} \url{https://en.wikipedia.org/wiki/Point_in_polygon/}
 \bibitem{Boris} J. Boris, in Proceedings of the Fourth International Conference on Numerical Simulation of Plasmas (NRL, 1970), p. 367.
 \bibitem{Birdsall} C. K. Birdsall and A. B. Langdon, Plasma Physics via Computer Simulation (Institute of Physics, New York, 2005).
  \bibitem{Tajima} T. Tajima, Computational Plasma Physics with Applications to Fusion and Astrophysics (Perseus, Boulder, 2004).
 \bibitem{wei2015method} X. Wei, Y. Xiao, A. Kuley, and Z. Lin, \href{https://doi.org/10.1063/1.4929799}{Phys. Plasmas} \textbf{22}, 092502 (2015). 
 \bibitem{brizard2007foundations} A. Brizard and T. Hahm, \href{https://doi.org/10.1103/RevModPhys.79.421}{Rev. of Modern Phys.} \textbf{79}, 421 (2007).

 
 \end{thebibliography}
\end{document}